\documentclass[prl,english,twocolumn,amsmath,amssymb,superscriptaddress]{revtex4-1}
\usepackage{epsfig, graphicx,graphics,amsmath,amssymb,float}
\usepackage[T1]{fontenc}
\usepackage[latin9]{inputenc}
\usepackage{amsmath}
\usepackage{amssymb}

\usepackage{amscd}
\usepackage{bm}
\usepackage{psfrag}
\usepackage{bbm} 
\usepackage{babel}
\usepackage{wasysym }
\usepackage{mathrsfs}
\usepackage{color}

\newcommand{\be}{\begin{equation}}
\newcommand{\ee}{\end{equation}}
\newcommand{\bea}{\begin{eqnarray}}
\newcommand{\eea}{\end{eqnarray}}

\newcommand{\mc}{\mathcal}
\newcommand{\mb}{\mathbf}

\begin{document}

\title{Electrical Access to Ising Anyons in Kitaev Spin Liquids}

\author{Rodrigo G. Pereira}
\affiliation{International Institute of Physics and Departamento de F\'isica Te\'orica
e Experimental, Universidade Federal do Rio Grande do Norte, 
Natal, RN, 59078-970, Brazil}
\author{Reinhold Egger}
\affiliation{Institut f\"ur Theoretische Physik,
Heinrich-Heine-Universit\"at, D-40225  D\"usseldorf, Germany}

\begin{abstract}
We show that spin-spin correlations in a non-Abelian Kitaev spin liquid are associated with a characteristic inhomogeneous charge density distribution 
in the vicinity of $\mathbb{Z}_2$ vortices.  This density profile and the corresponding local electric fields are observable, e.g.,
by means of surface probe techniques.  Conversely, by applying bias voltages to several probe tips, one can stabilize Ising anyons 
($\mathbb{Z}_2$ vortices harboring a Majorana zero mode) at designated positions, where we predict a clear Majorana signature 
in energy absorption spectroscopy.  
\end{abstract}
\maketitle

\emph{Introduction.---}Quantum spin liquids (QSLs) are fascinating topologically ordered  phases of  
quantum spins on lattices with frustrated interactions \cite{Savary2017,Zhou2017,Wen2017,Knolle2019,Broholm2020}.   
Kitaev's celebrated two-dimensional (2D) honeycomb model in a magnetic field \cite{Kitaev2006}  provides an exactly solvable example for a non-Abelian chiral spin liquid, 
featuring emergent gapped neutral fermions as well as Ising anyons --- Majorana zero modes (MZMs) bound to $\mathbb{Z}_2$ vortices --- as elementary bulk excitations. In addition, a gapless chiral Majorana fermion mode at the boundary is responsible for a quantized thermal Hall effect.  The Kitaev model can be approximately realized in different material platforms \cite{Jackeli2009,Winter2017,Trebst2017,Hermanns2018,Takagi2019}, e.g., in $\alpha$-RuCl$_3$ \cite{Plumb2014,Sandilands2015,Banerjee2016} where small inter-layer couplings indicate that 2D models are appropriate \cite{Kim2015,Kim2016}. 
Recent experiments suggest a Kitaev spin liquid phase in $\alpha$-RuCl$_3$ at intermediate magnetic field strength between a magnetically ordered
low-field state and a polarized high-field phase \cite{Baek2017,Sears2017,Wolter2017,Leahy2017,Banerjee2018,Hentrich2018,Balz2019,Tanaka2020}. In particular, the thermal Hall signature of the chiral Majorana edge mode has been reported \cite{Kasahara2018,Yokoi2020,Motome2020}, see also Refs.~\cite{Vinkler2018,Ye2018}. Nonetheless, the question of whether a QSL phase has really been observed in $\alpha$-RuCl$_3$ remains controversial, see, e.g., Refs.~\cite{Sahasrabudhe2020,Bachus2020}.  

It stands to reason that alternative experimental techniques can help to unambiguously identify QSL physics in Kitaev materials. 
Recent theoretical works \cite{Aasen2020,Zhang2020} have suggested \emph{electrical} detection methods --- even though QSLs are realized in charge insulating magnetic materials.
 Aasen \textit{et al.}~\cite{Aasen2020} (see also Ref.~\cite{Barkeshli2014}) argue that Ising anyons and/or the chiral Majorana mode can be detected
by measuring the electrical conductance in circuits where quantum Hall edges and superconductors are strongly coupled to the QSL. 
We here describe a different but also purely electrical approach for detecting and manipulating Ising anyons in  Kitaev spin liquids.  
Noting that vacancies or magnetic impurities allow to trap $\mathbb{Z}_2$ vortices \cite{Dhochak2010,Willans2011,Vojta2016}, 
our ideas may guide efforts towards establishing Kitaev materials as useful platform for topological quantum information processing \cite{Kitaev2006,Aasen2020,Nayak2008}.

Our work is motivated by the fact that Mott insulators can harbor quantum states with nontrivial electric polarization profile \cite{Katsura2005,Katsura2007,Bulaevskii2008,Khomskii2010}. Similarly, spin excitations in QSLs may contribute to the optical conductivity inside the Mott gap \cite{Ng2007,Potter2013,Little2017,Bolens2018}.  
For instance,  consider a half-filled Hubbard model on an arbitrary 2D lattice at strong coupling, $|t_{jk}|\ll U$, with 
on-site interaction $U$ and (real) tunnel couplings $t_{jk}$ between sites $j$ and $k$. 
Writing the electron density operator at site $j$ as $\hat n_j=1+\delta \hat n_j$, one finds that $\delta \hat n_j=\sum_{k,l}\delta \hat n_{j,(kl)}$ can be expressed
by the low-energy spin-$1/2$ operators ${\mb S}_j=(S^x_j,S^y_j,S^z_j)$ \cite{Bulaevskii2008,Khomskii2010},
\be\label{deltan}
 \delta\hat n_{j,(kl)} = \frac{8t_{jk}t_{kl}t_{lj}}{U^3}\left(\mb S_j\cdot \mb S_k+ \mb S_j\cdot \mb S_l-2\mb S_k\cdot \mb S_l\right ).
\ee 
The ground state (g.s.) charge imbalance at site $j$ follows by summing the spin-spin correlations over all triangular site
configurations $(jkl)$. While Eq.~\eqref{deltan} implies overall charge neutrality, $\sum_{j}\delta\hat n_j=0$, inhomogeneous charge densities emerge for spin correlations with nontrivial spatial structure. The physical intuition is that electronic charge can be locally attracted to (or repelled by) a strong exchange bond, 
depending on the signs of spin correlations and tunnel couplings \cite{Bulaevskii2008,Khomskii2010}.  

Previous works have examined such phenomena in the context of noncollinear magnetism \cite{Katsura2005,Katsura2007,Bulaevskii2008,Khomskii2010}. We here 
study the local charge imbalance in a Kitaev QSL harboring Ising anyons, where the spin-SU$(2)$ symmetric result \eqref{deltan} does not apply.
Starting from a multi-orbital Hubbard-Kanamori model \cite{Jackeli2009}, 
 the polarization profile again follows by summing certain spin correlations over triangular site configurations, see Eq.~\eqref{deltan2} below.
Exploiting that spin correlations can be calculated in an exact manner for arbitrary eigenstates of the 
Kitaev model \cite{Baskaran2007,Knolle2014}, we demonstrate that a $\mathbb{Z}_2$ vortex will induce a radially symmetric and oscillatory charge density profile. 
Using surface probe techniques like atomic force microscopy (AFM) or scanning tunneling microscopy (STM) for 
exfoliated/cleaved $\alpha$-RuCl$_3$ samples \cite{Ziatdinov2016,Weber2016,Du2018,Feldmeier2020}, this profile and the associated electric 
fields could be detected experimentally.  Similarly, by applying a voltage to a probe tip, 
$\mathbb{Z}_2$ vortices can be stabilized below the tip. Upon slowly moving the tip in the lateral direction, the vortex may then be transported
to a designated position.  Finally, in a setup with four tips, the existence of MZMs may be verified by energy absorption spectroscopy 
\cite{Rugar2004,Baumann2015}. 

\begin{figure}[t]
\begin{center}
\includegraphics[width=\columnwidth]{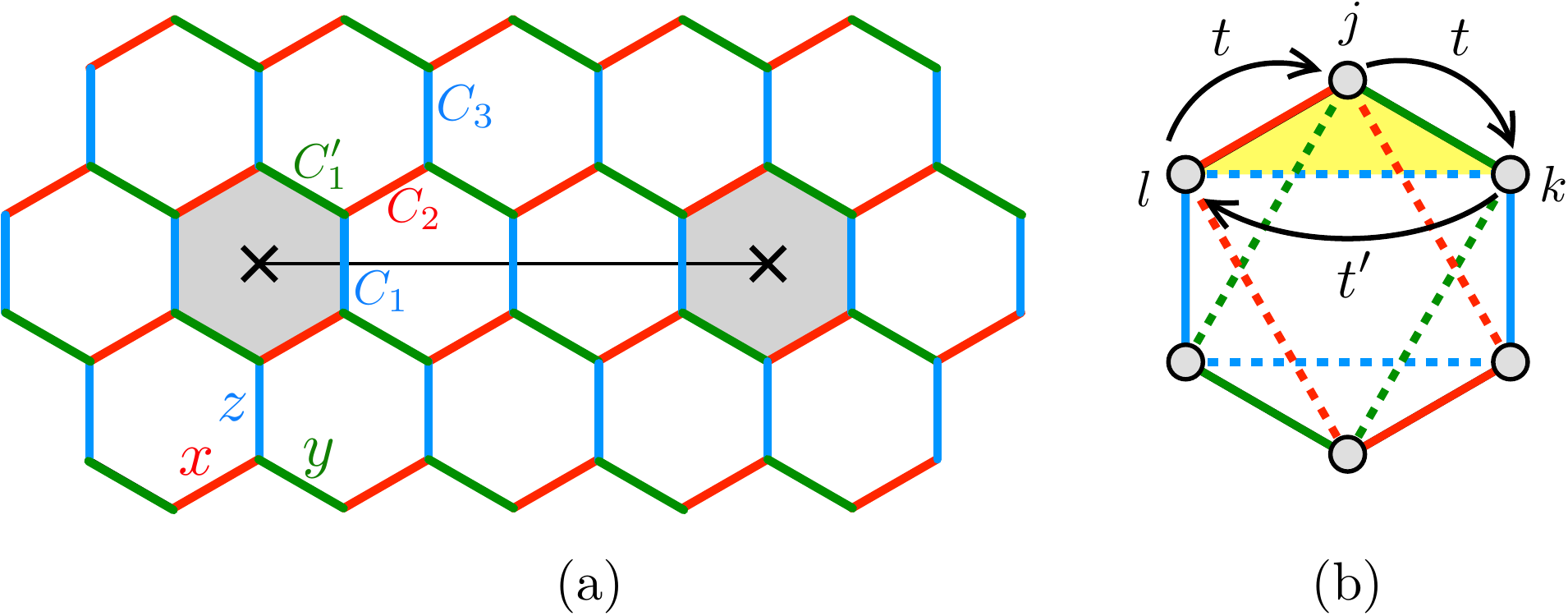}
\end{center}
\caption{(a) Two-vortex state in the Kitaev model, where filled hexagons represent $\mathbb{Z}_2$ vortices with $W_p=-1$. Compared to the uniform g.s.~with all $W_p=+1$,
the gauge fields $u_{\langle jk\rangle_\alpha}$ are reversed for bonds crossed by a string (solid black line) connecting both vortices. 
Different bond types, $\alpha=(x,y,z)$, are depicted in red, green and blue, respectively. 
Spin-spin correlations, see Eq.~\eqref{gjl}, near a vortex become spatially isotropic in the thermodynamic limit and with all other vortices far away, i.e., 
$C_1'=C_1$. 
(b) A local charge imbalance $\rho_j$ arises from virtual orbital-dependent hopping around triangles with one bond of each type $\alpha$, with amplitude $t$ for nearest-neighbor bonds (solid lines) and $t'$ for next-nearest-neighbor bonds (dashed lines). For the site with index $j$, the yellow triangle gives a contribution $\rho_j\propto 2C_1$. 
}
\label{fig1}
\end{figure}

\emph{Kitaev model.---}Consider the exactly solvable Kitaev honeycomb model with symmetric exchange couplings
in a weak magnetic field ${\bf h}$ \cite{Kitaev2006},
\be\label{kappa}
H = - K \sum_{\langle jl\rangle_\alpha} S_j^\alpha S_l^\alpha -\kappa\sum_{\langle jk\rangle_\alpha ,\langle kl\rangle_\beta} 
S_j^\alpha  S^\gamma_k S_l^\beta,
\ee
where $\langle jl\rangle_\alpha$ denotes a nearest-neighbor bond of type $\alpha=x,y,z$,  see Fig.~\ref{fig1}. The term $\propto \kappa$ 
encodes the magnetic field, where $(\alpha\beta\gamma)$ is a cyclic permutation of $(xyz)$ and the sum extends over triangles $(jkl)$ with two adjacent
nearest-neighbor bonds.
While a perturbative calculation yields $\kappa \propto h_xh_yh_z/K^2$ \cite{Kitaev2006}, in more general models beyond Eq.~\eqref{kappa},
$\kappa$ is already generated at first order in $|{\bf h}|$ \cite{Song2016}.  Throughout we assume $\kappa\ne 0$ and measure lengths in lattice spacing ($a_0$) units, where $a_0=3.44$~\AA~for $\alpha$-RuCl$_3$ \cite{Kim2016}. 

The model \eqref{kappa} is diagonalized by using a Majorana representation of the spin-$1/2$ operators,
$S_j^\alpha =\frac{i}{2} c^{}_j c_j^\alpha$, with anticommuting Majorana operators $\left(c_j^{},c_j^\alpha\right)$ squaring to unity \cite{Kitaev2006}.
One first defines $\mathbb Z_2$ gauge fields, $u_{\langle jl\rangle_\alpha}=ic^\alpha_jc^\alpha_l=-u_{\langle lj\rangle_\alpha}$, which are conserved bond operators with eigenvalue $\pm1$.
For given gauge field configuration $|\mc G\rangle$, Eq.~\eqref{kappa} describes noninteracting Majorana fermions $\{ c_j\}$,
\be\label{HMajorana}
H=\frac{iK}{4}\sum_{\langle jl\rangle_\alpha} u_{\langle jl\rangle_\alpha}c_jc_l-\frac{i\kappa}{8} \sum_{\langle jk\rangle_\alpha, \langle kl\rangle_\beta}u_{\langle jk\rangle_\alpha}u_{\langle kl\rangle_\beta}c_jc_l,
\ee
with eigenstates $|\mc M(\mc G)\rangle$.
Clearly, spin operators are invariant under $\mathbb Z_2$ gauge transformations, $\left(c^{}_j,c_j^\alpha\right)\mapsto \left(-c^{}_j,-c_j^\alpha\right)$. 
Since the gauge structure artificially enlarges the Hilbert space, the local constraint $D_j\equiv c^{}_jc_j^xc_j^yc_j^z=1$ is imposed by
the projector $\mc P=\prod_j\frac{1+D_j}{2}$, and  the exact eigenstates are given by $|\Psi\rangle=\mc P|\mc M(\mc G)\rangle\otimes |\mc G\rangle$.
The gauge invariant $\mathbb{Z}_2$ flux through the $p$th hexagon is encoded by the plaquette operator
$W_p=\prod_{\langle jl\rangle_\alpha\in \varhexagon_p} u_{\langle jl\rangle_\alpha}=\pm 1$ (with bonds oriented from $j$ in sublattice A to $l\in$~B), where
 $W_p=-1$ defines a $\mathbb{Z}_2$ vortex, see Fig.~\ref{fig1}(a).
The g.s.~sector has no vortices and is solved by 
Fourier transformation of Eq.~\eqref{HMajorana} with all $u_{\langle jl\rangle_\alpha }\to +1$ \cite{Kitaev2006}.

For arbitrary eigenstates $|\Psi\rangle$, spin correlations can be computed in an exact manner.  
They vanish except for nearest-neighbor bonds $\langle jl\rangle_\alpha$, where one finds \cite{Baskaran2007,Knolle2014}
\begin{eqnarray}\label{gjl}
\langle\Psi| S^\alpha_j S^\beta_l|\Psi\rangle &= &\frac{1}{4}C_{\langle jl\rangle_\alpha}\delta^{\alpha\beta},\\ \nonumber
C_{\langle jl\rangle_\alpha}&=&-  u_{\langle jl\rangle_\alpha}   \langle \mc M(\mc G )|i c_j c_l|\mc M(\mc G )\rangle.
\end{eqnarray}
This ultralocal behavior is due to the static nature of the gauge field --- vortices created by $S_l^\beta$
must be annihilated by $S^\alpha_j$ again.  Details on the numerical calculation of $C_{\langle jl\rangle_\alpha}$ 
in a finite Kitaev lattice with $\mathbb{Z}_2$ vortices are provided in the Supplementary Material (SM) \cite{SM}.

\emph{Charge density in Kitaev materials.---}Kitaev materials correspond to multi-orbital Mott insulators with strong spin-orbit coupling \cite{Jackeli2009}. We now generalize Eq.~\eqref{deltan} 
and relate the local density operator, $\hat n_j=1+\delta\hat n_j$, to spin-spin correlations in Kitaev materials. 
We start from the  Hubbard-Kanamori model for $d^5$ electrons in an  edge-sharing octahedral environment 
\cite{Jackeli2009,Rau2014,Rau2016,Winter2016}, with on-site Coulomb energy $U$, Hund coupling $J_H=\eta U$ (with $0<\eta<1/3$), and real-valued positive hopping 
amplitudes $t$ and $t'$, see Fig.~\ref{fig1}(b).  We consider only the dominant hopping path which couples, e.g., $xz$ and $yz$ orbitals on $z$ bonds 
\cite{Jackeli2009}.  Assuming $t,t'\ll U$,  a canonical transformation \cite{Takahashi1977,MacDonald1988,Chernyshev2004}
projects this model to the low-energy sector, where the single hole at each site is in a state with effective total angular momentum $j_{\text{eff}}=1/2$.
With ${\bf S}_j$ now referring to hole spin-$1/2$ operators, one arrives at the Kitaev model \cite{Jackeli2009}  with  
 $K=\frac{8\eta t^2}{3 (1-\eta)(1-3\eta) U }>0$, 
plus next-nearest neighbor Kitaev couplings $\propto (t^\prime)^2/U$. 
Performing the canonical transformation at next order in $(t,t')/U$ \cite{SM}, 
the local charge operator follows as $\delta\hat n_j=\sum_{k,l} \delta \hat n_{j,(kl)}$, summed over all 
triangular configurations with bond type $(\alpha,\beta,\gamma)$ of pair $(jk,jl,kl)$, respectively,
\bea\nonumber
\delta \hat n_{j,(kl)}& =& 
A_1S_k^\alpha S_j^\alpha + A_2 S_k^\beta S_j^\beta + A_3S_k^\gamma S_j^\gamma \\  \label{deltan2}
&+& A_1 S_l^\beta S_j^\beta + A_2 S_l^\alpha S_j^\alpha +A_3 S_l^\gamma S_j^\gamma \\
&-& \nonumber
2 A_1S_k^\gamma S_l^\gamma - (A_2+A_3) \left(S_k^\alpha S_l^\alpha +  S_k^\beta S_l^\beta\right) ,
\eea
with 
\be \label{adef}
\left(\begin{array}{c} A_1\\
A_2\\ A_3 \end{array}\right) =\frac{4\eta t^2 t^\prime}{9(1-\eta)^3(1-3\eta)^3U^3}
 \left(\begin{array}{c}
3-10 \eta+11 \eta ^2\\
5-20 \eta+21 \eta^2 \\
-5+18\eta-17\eta^2\end{array}\right).
\ee
Employing \emph{ab initio} parameters for $\alpha$-RuCl$_3$ \cite{Winter2017,Winter2016},
\be\label{parameters}
t= 160~{\rm meV},\  t'= 60~{\rm meV}, \ J_H= 0.4~{\rm eV}, \ U= 2.4~{\rm eV},
\ee
we obtain, e.g., $A_1\simeq 1.86\times 10^{-4}$.
We note that $\delta\hat n_j\propto J_H$, see Eq.~\eqref{adef}, suggests that the 
interference mechanism described in Ref.~\cite{Jackeli2009} also determines the electric polarization.

We here neglect additional interactions beyond the 
Kitaev couplings as well as subleading magnetic-field contributions to Eq.~\eqref{deltan2}~\cite{SM}.
To leading order in $t'/U$, the local charge imbalance, $\rho_j \equiv e\langle\Psi|\delta\hat n_j|\Psi\rangle$,
is then obtained from Eqs.~\eqref{deltan2} and \eqref{adef} by using the eigenstates $|\Psi\rangle$ of the pure ($t'=0$) Kitaev model.  
Employing the spin correlations in Eq.~\eqref{gjl} and summing over all triangles $(jkl)$, 
\be\label{honeycomb}
\rho_j = eA_1 \sum_{kl} \left( \big\langle S_k^\alpha S_j^\alpha\big\rangle+
\bigl\langle S_l^\beta S_j^\beta\bigr\rangle-2\big\langle S_k^\gamma S_l^\gamma\big\rangle \right).
\ee
For the uniform g.s.~without vortices, one readily shows $\rho_j=0$.  

\emph{Charge density near a vortex.---}Let us now consider a gauge state $|\mc G\rangle$ with two vortices 
at distance $d$, see Fig.~\ref{fig1}(a).  
We numerically diagonalize the Hamiltonian \eqref{HMajorana} for system size $L\times L$ with 
periodic boundary conditions, where spin correlations follow from Eq.~\eqref{gjl} \cite{SM}. 
The MZM operators $\gamma_{1,2}$ for both vortices allow for two parity 
states, $|n_f=0,1\rangle$ with $i\gamma_1\gamma_2=(-1)^{n_f}$. However, the total number of fermions is
subject to a parity constraint \cite{Pedrocchi2011,Zschocke2015} which fixes $n_f$, and hence
more than two vortices are needed for implementing nontrivial MZM operations in a Kitaev QSL.
Next we note that the string in Fig.~\ref{fig1} is gauge dependent and thus unphysical \cite{Savary2017,Wen2017}. 
With other vortices far away, spin correlations near a $\mathbb{Z}_2$ vortex must therefore become isotropic, e.g., $C_1^\prime=C_1$ in Fig.~\ref{fig1}(a).
From Eq.~\eqref{honeycomb}, the local charge imbalance then depends on at most three non-equivalent spin correlators $C_{1,2,3}$, see
 Fig.~\ref{fig1}(a) for sites surrounding a vortex. Summing over all triangles, we find $\rho_j =   eA_1  (C_1-C_3)$ at those sites, where $C_2$ cancels out identically. 
We have numerically computed $C_1-C_3$ for the largest inter-vortex distance $d=\lfloor (L-1)/2 \rfloor$ \cite{SM}. For $\kappa=0.2K$,   
the thermodynamic limit is reached for $L\agt 20$. We then find $C_1-C_3\simeq -0.0315$, resulting in $\rho_j\simeq -5.86 \times 10^{-6}e$.

\begin{figure}
\begin{center}
\includegraphics[width=\columnwidth]{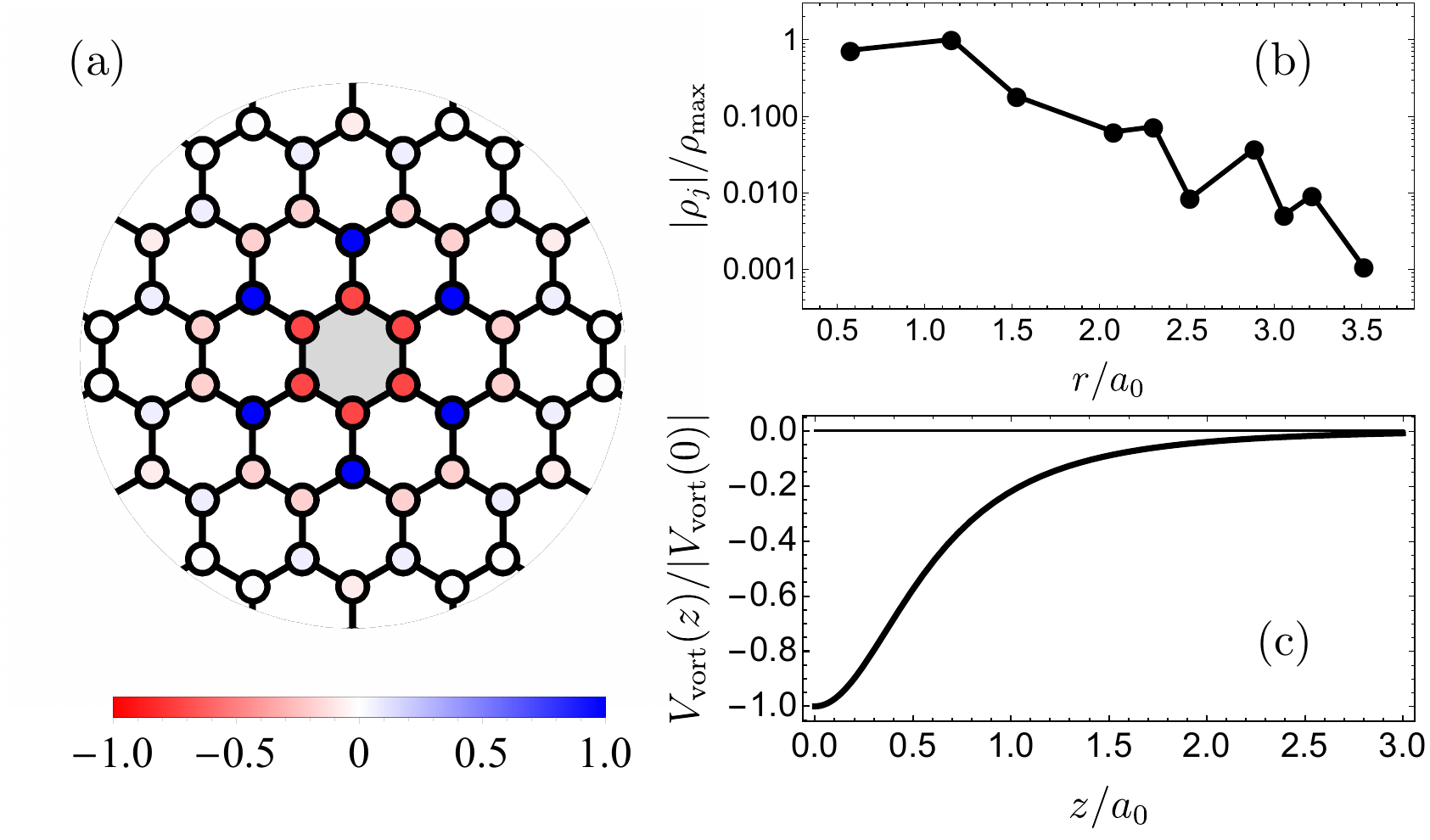}
\end{center}
\caption{Charge density profile, $\rho_j$, near a vortex (shaded hexagon) and the corresponding electrostatic potential, $V_{\rm vort}(z)$, for $\kappa=0.2K$ and parameters \eqref{parameters}. (a) Color-scale plot of $\rho_j/\rho_{\rm max}$, see Eq.~\eqref{honeycomb}.  
(b) $|\rho_j| /\rho_{\rm max}$ vs in-plane distance $r$ from the vortex center on a semi-logarithmic scale. 
(c) $V_{\rm vort}(z)$ vs perpendicular distance $z$, with $V_{\rm vort}(z=0)\simeq -0.118$~meV at the vortex center. }
\label{fig2}
\end{figure}

The full charge density profile around a single vortex (with all other vortices far away) is shown in Fig.~\ref{fig2}(a). 
The profile is radially isotropic and exhibits Friedel-like oscillations with
the distance $r$ from the vortex center, where the largest charge imbalance, $\rho_{\rm max}\equiv {\rm max}|\rho_j|\simeq 8.09\times 10^{-6}e$, occurs at the second `ring'.  Moreover, Fig.~\ref{fig2}(b) indicates exponentially small charge imbalances for large $r$. 

\emph{Vortex detection.---}A $\mathbb{Z}_2$ vortex can be detected through the associated charge density profile in STM measurements \cite{Ziatdinov2016}, or 
by mapping out the resulting local electric fields, e.g., using AFM techniques \cite{APL,Wagner2019,Mohn2012}. 
The electrostatic potential at position ${\mb r}$ follows from Eq.~\eqref{honeycomb} by summing over all honeycomb lattice sites $\mb R_j$, $V_{\rm vort}(\mb r)= \sum_j \rho_j/|\mb r-\mb R_j|$,  see Fig.~\ref{fig2}(c).  
This polarization profile generates a quadrupole potential which is most pronounced along the perpendicular direction.
Putting $\mb r=z\hat e_z$, the numerical results in Fig.~\ref{fig2}(c) are consistent with $V_{\rm vort}(z)\propto -1/z^3$ for $|z|\to \infty$.
Since available AFM techniques resolve voltage differences far below $0.1$~mV \cite{APL,Wagner2019,Mohn2012}, experimental tests of this prediction are within reach.

\emph{Vortex manipulation.---}We next turn to the influence of local external electric fields. For definiteness, 
we consider a voltage-biased (AFM or STM) probe tip positioned above a hexagon center.  We approximate the tip potential 
by a constant, $V_0$, for all six sites around the hexagon, and zero otherwise. Including the electrostatic coupling in the 
atomic on-site term of the Hubbard-Kanamori model, we again project to the low-energy sector with a single $j_{\rm eff}=1/2$ hole per site \cite{SM}.
Using $\xi_0\equiv eV_0/[(1-3\eta)U]$, this projection applies for $|\xi_0|<1$.
We arrive at the Kitaev model \eqref{kappa}, where the exchange couplings for the $C_2$-bonds in Fig.~\ref{fig1}(a) are 
instead of $K$ given by
\be
K(V_0)= \frac{(1 +\nu_0 \xi_0^2)\,K}{(1-\xi_0^2)(1- \nu_0^2\xi_0^2)}\ge K , \quad \nu_0=\frac{1-3\eta}{1-\eta}.\label{KwithV0}
\ee
With increasing $|V_0|$, the exchange coupling also increases. (For simplicity, we assume $\kappa(V_0)=\kappa$.)

\begin{figure}
\begin{center}
\includegraphics[width=\columnwidth]{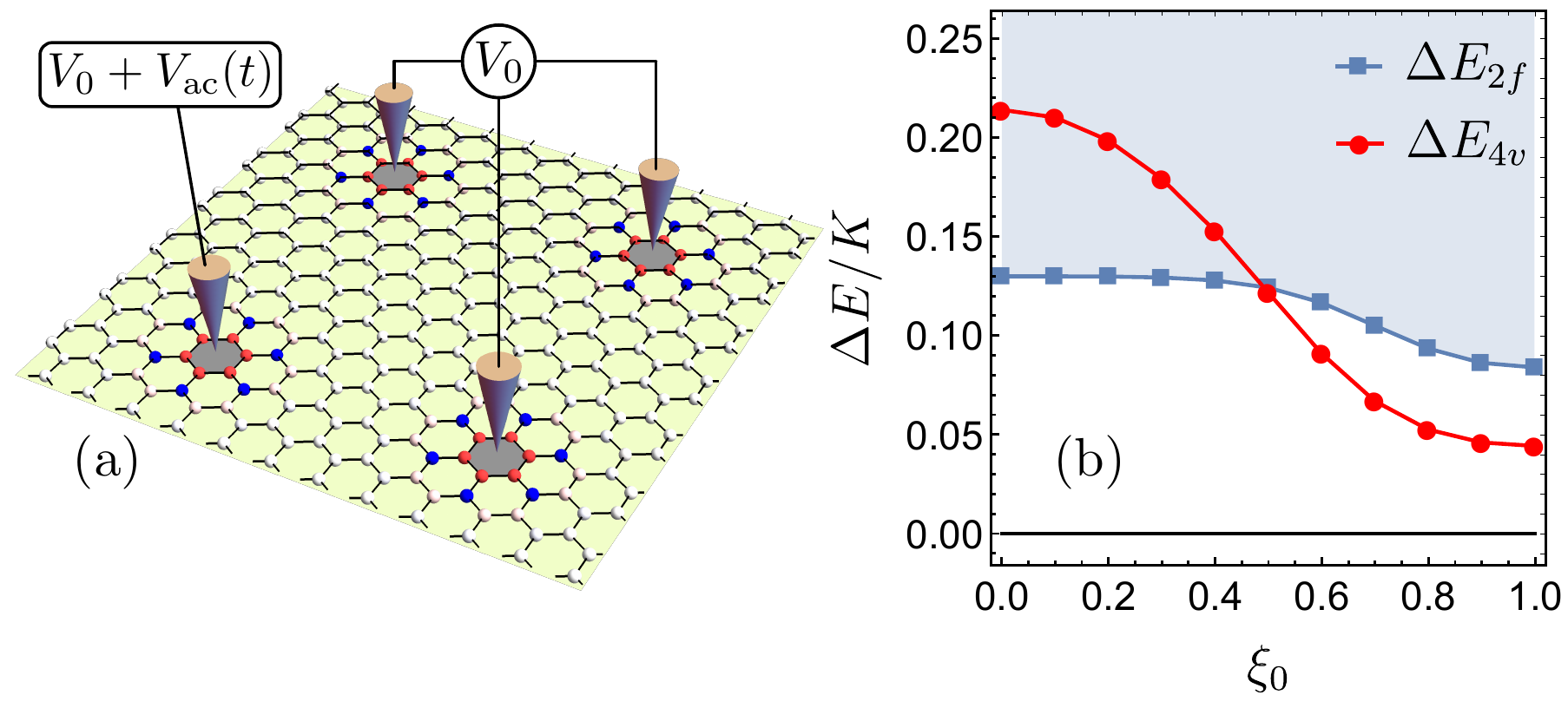}
\end{center}
\caption{(a) Setup with four probe tips held at the same voltage $V_0$. 
By applying a weak a.c.~voltage $V_{\rm ac}(t)$ to one tip,
the energy absorption probability \eqref{Pomega} can be measured.  This quantity 
provides information about the low-energy QSL excitation spectrum.
(b) Energy gap of the lowest four-vortex state, $\Delta E_{4v}$ (red), and two-fermion state, $\Delta E_{2f}$ (blue), vs 
tip voltage parameter $\xi_0=eV_0/[(1-3\eta)U]$.  We use the parameters in Eq.~\eqref{parameters}, $\kappa=0.1K$, $V_{\rm ac}=0$, $L=30$, with all tips far away from each other. The shaded region shows the two-fermion continuum 
without vortices.  }
\label{fig3}
\end{figure}

We then consider a setup with four tips at the same voltage $V_0$, see Fig.~\ref{fig3}(a) for $V_{\rm ac}=0$.
The resulting Kitaev model remains exactly solvable since only the $C_2$-bonds around each of the four hexagons are modified.
Figure \ref{fig3}(b) shows the energy gap $\Delta E_{4v}$ from the uniform g.s.~without vortices to the g.s.~with four vortices at the contacted hexagons, as well as the gap to the first excited state without vortices, $\Delta E_{2f}$, where two bulk fermions are created. 
With increasing $V_0$, we observe that $\Delta E_{4v}$ decreases and eventually falls below $\Delta E_{2f}$.  
A vortex located near one of the probe tips will thus be \emph{attracted} towards the position right below the tip.
For this voltage-controlled trapping mechanism, using $K\approx 5$~meV \cite{Winter2017}, $\xi_0=0.5$, and the parameters in Fig.~\ref{fig3}(b), the stabilization energy is $\Delta E_{4v}(0)-\Delta E_{4v}(V_0) \approx 0.5$~meV.  Once a vortex has been trapped, by slowly dragging the probe tip along the lateral direction, one could 
transport the vortex to a desired position.  

Since $V_0\ne 0$ breaks the symmetry between sites on a bond, a local charge imbalance is already possible for $t'=0$.  
Writing $\delta \hat n_j=\sum_l \delta \hat n^{(2)}_{j,l}$, where one sums over bonds $\langle jl\rangle_\alpha$, we find 
\be\label{deltanV0}
\delta \hat n_{j,l}^{(2)}=-\frac{4t^2eV_0}{U^3}\left[f_0\left(\xi_0,\eta\right)+f_s\left(\xi_0,\eta\right)  S^\alpha_j  S^\alpha_l\right].
\ee
For $\xi_0\ll 1$ and $\eta\ll 1$, the functions $f_{0,s}\left(\xi_0,\eta\right)$ approach 
$f_0\simeq 2/3$ and $f_s\simeq 4\eta$ \cite{SM}. The superscript `(2)' indicates that   
second-order contributions now dominate over the third-order terms in Eq.~\eqref{deltan2}. 

\begin{figure}
\begin{center}
\includegraphics[width=\columnwidth]{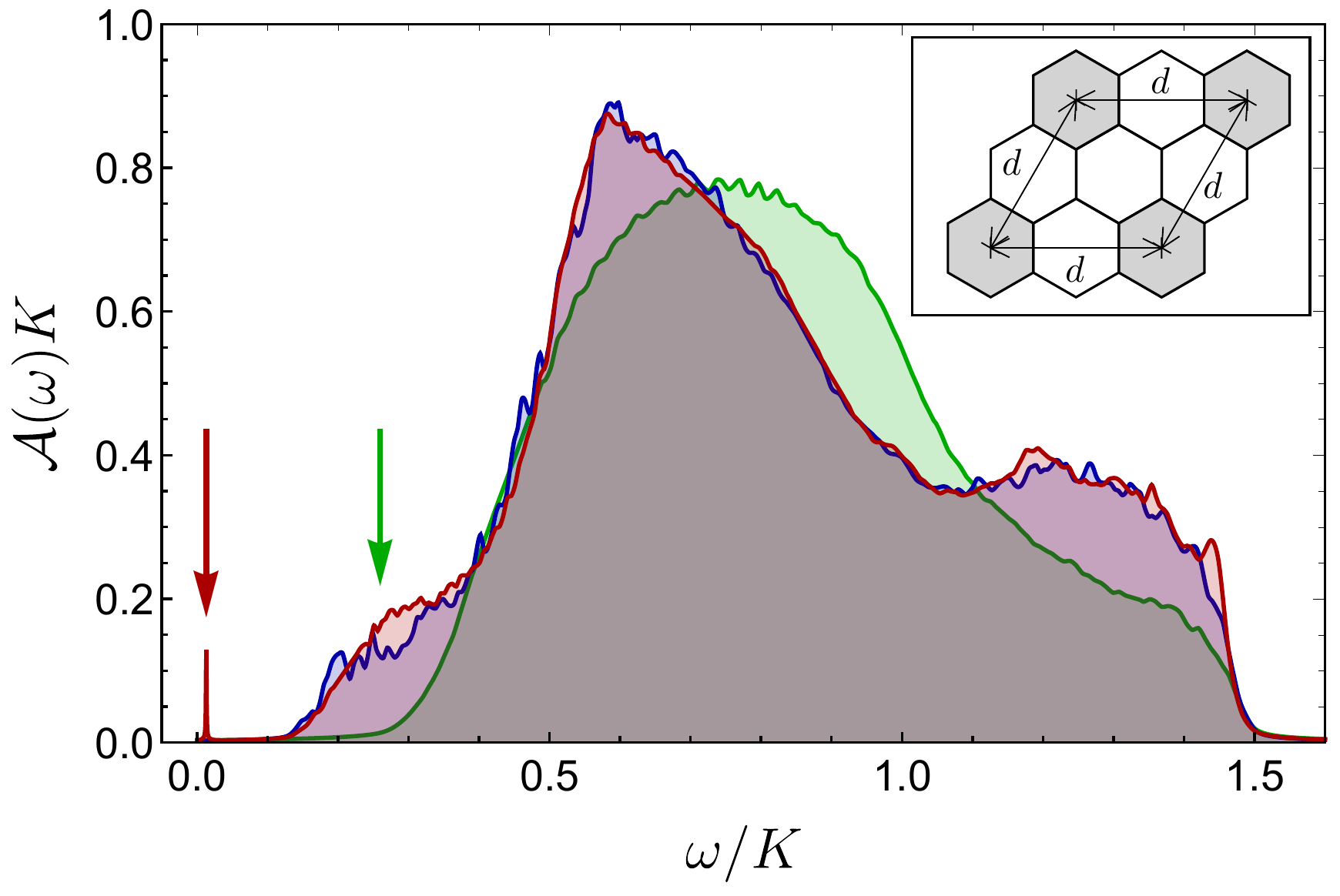}
\end{center}
\caption{Line shape $\mc A(\omega)$ of the absorption spectrum \eqref{Pomega} for the setup in Fig.~\ref{fig3}(a). We use the parameters \eqref{parameters},
$\xi_0=0.2$, $\kappa=0.2K$, $L=48$, $d=2$ (see inset), and different numbers $N_V$ of vortices below the tips: 
$N_V=0$ (green), $N_V=2$ (blue, including a vortex at the first tip), and $N_V=4$ (red curve). The green arrow indicates the $N_V=0$ two-fermion continuum threshold $\omega=\Delta E_{2f}$.
The red arrow marks the MZM peak at $\omega=\varepsilon_M(d)$, possible only for $N_V=4$.  
For the continuum part, $\delta$-functions are broadened by the maximal level spacing, $(\delta \omega)_{N_V=0}\approx 8.7\times 10^{-3}K$ and 
$(\delta \omega)_{N_V=2,4}\approx 5.5\times 10^{-3}K$. For the MZM peak, $\delta\omega= 5\times 10^{-4}K$. }
\label{fig4}
\end{figure}

\emph{Energy absorption spectroscopy.---}Finally, we outline a spectroscopic technique for detecting the MZMs bound by $\mathbb{Z}_2$ vortices.  We consider a four-tip setup with nearest-neighbor tip distance $d$ and all tips at the same voltage $V_0$, see Fig.~\ref{fig3}(a).  Accounting for the total fermion parity constraint \cite{Pedrocchi2011,Zschocke2015}, MZMs cause a two-fold g.s.~degeneracy for $N_V=4$ vortices and $d\to \infty$.  At finite $d$, an exponentially small energy splitting, $\varepsilon_M(d)$, is present \cite{Kitaev2006}.
On the first tip, we add a weak a.c.~voltage, $V_{\rm ac}(t)=V_1\cos(\omega t)$ with  $V_1\ll V_0$.   
The energy absorption probability follows from Fermi's golden rule,
\bea\label{Pomega}
P(\omega)&=&\left(\frac{2Ke^2V_0V_1w\left (\xi_0,\eta\right)}{(1-3\eta)^2 U^2}\right)^2\mc A(\omega),\\ \nonumber
\mc A(\omega)&=& 2\pi\sum_{n} |\langle \Psi_n |\hat Q_1|\Psi_0\rangle|^2 \delta(\omega -E_n+E_0),
\eea
with unperturbed ($V_1=0$) eigenstates $|\Psi_n\rangle$ for energy $E_n$, where the g.s.~corresponds to $n=0$. The function $w(\xi_0,\eta)$ \cite{SM} approaches $w\simeq 3$ for $\xi_0\ll 1$ and $\eta\ll 1$. Noting that $V_{\rm ac}(t)$ couples to the charge accumulated below the first tip, $\hat Q_1$ corresponds to the two-spin operator 
 \be
 \hat Q_1= \sum_{\langle jl\rangle_\alpha} S_j^\alpha S_l^\alpha =-\frac{i}{4}\sum_{\langle jl\rangle_\alpha} u_{\langle jl\rangle_{\alpha}}c_jc_l,
 \ee
 where one sums over $C_2$-bonds at the first hexagon.
Matrix elements of $\hat Q_1$ can couple the g.s.~to excited states without changing $N_V$, in contrast to single-spin operators \cite{Knolle2014}. Importantly, $P(\omega)$ can be measured by spectroscopic techniques as introduced in Refs.~\cite{Rugar2004,Baumann2015}.

Our results for the absorption spectrum \eqref{Pomega} are shown in Fig.~\ref{fig4}.  
For $N_V=4$, a sharp MZM peak at $\omega=\varepsilon_M(d)$ emerges well below the continuum part, where the $N_V=2,4$ continuum threshold involves one bulk fermion and a zero mode, i.e.,
$\omega\approx\Delta E_f$. Since $\Delta E_f<\Delta E_{2f}$, see Fig.~\ref{fig4}, this $N_V$-dependence of the 
continuum threshold  would give direct evidence for MZMs. For $d\to \infty$, the MZM peak weight vanishes because the local operator $\hat Q_1$ cannot distinguish degenerate topological ground states.  
By monitoring the $d$-dependence of this peak, however, MZMs could also be detected.

\emph{Conclusions.---}The inhomogeneous charge density near $\mathbb{Z}_2$ vortices allows one to detect and manipulate Ising anyons via local electric fields, where surface probe techniques
could eventually enable MZM fusion and braiding \cite{Kitaev2006,Nayak2008} experiments in Kitaev spin liquids.
As robust and feasible prediction, the electric polarization profile and the energy absorption spectrum must change when a vortex is trapped at (or 
removed from) a plaquette near an STM tip.

\begin{acknowledgments}
We acknowledge funding by the Brazilian ministries MEC and MCTI, by the Brazilian agency CNPq, and by the Deutsche Forschungsgemeinschaft (DFG, German Research Foundation),
Projektnummer 277101999 - TRR 183 (project B04) and under Germany's Excellence Strategy - Cluster of Excellence Matter
and Light for Quantum Computing (ML4Q) EXC 2004/1 - 390534769.
\end{acknowledgments}

\onecolumngrid

\appendix
 
\section{Supplemental Material to ``Electrical Access to Ising Anyons in Kitaev Spin Liquids''}

\author{Rodrigo G. Pereira}
\affiliation{International Institute of Physics,  Universidade Federal do Rio Grande do Norte, 
Natal, RN, 59078-970, Brazil}
\affiliation{Departamento de F\'isica Te\'orica
e Experimental, Universidade Federal do Rio Grande do Norte, 
Natal, RN, 59078-970, Brazil}
\author{Reinhold Egger}
\affiliation{Institut f\"ur Theoretische Physik,
Heinrich-Heine-Universit\"at, D-40225  D\"usseldorf, Germany}

\section{I. Spin correlations in the Kitaev model} \label{sec1}

We begin by describing the calculation of spin-spin correlations
using the Majorana representation of the Kitaev model with $\kappa\ne 0$, see Eqs.~(3) and (4) of the main text and Refs.~\cite{Kitaev2006,Baskaran2007}. 
We consider spin-$1/2$ operators on a 2D honeycomb lattice with $L\times L$ unit cells. The $2N=2L^2$ Majorana operators $c_j$ introduced in the main text
are written as $c_j=c_\lambda(m,n)$, with sublattice index $\lambda\in \{{\rm A,B}\}$
and integers $m,n=1,\dots,L$ labeling the unit cells, $\mb R(m,n)=m\hat{\mb e}_1+n\hat{\mb e}_2$. The
primitive lattice vectors are $\hat{\mb e}_1=\frac12\hat{\mb x}+\frac{\sqrt3}{2}\hat{\mb y}$ and  $\hat{\mb e}_2=-\frac12\hat{\mb x}+\frac{\sqrt3}2\hat{\mb y}$, and
we use periodic boundary conditions, $c_\lambda(m+L,n)=c_\lambda(m,n)$ and $c_\lambda(m,n+L)= c_\lambda(m,n).$
The Hamiltonian then reads
\begin{widetext}
\bea
H&=&i\frac{K}{4}\sum_{m,n}c_{\rm A}(m,n)\Bigl[u_z(m,n)c_{\rm B}(m,n)+u_x(m,n)c_{\rm B}(m+1,n)+u_y(m,n)c_{\rm B}(m,n+1)\Bigr] \nonumber\\
&&+i\frac{\kappa}{8} \sum_{m,n}\biggl\{c_{\rm A}(m,n)\Bigl[u_x(m,n)u_y(m+1,n-1)c_{\rm A}(m+1,n-1)
+u_z(m,n)u_x(m-1,n)c_{\rm A}(m-1,n)\nonumber\\ &&\nonumber +u_y(m,n)u_z(m,n+1)c_{\rm A}(m,n+1)\Bigr]  +c_{\rm B}(m,n)\Bigl [ u_x(m-1,n)u_y(m-1,n)c_{\rm B}(m-1,n+1)\\
&&+u_z(m,n)u_x(m,n)c_{\rm B}(m+1,n)+u_y(m,n-1)u_z(m,n-1)c_{\rm B}(m,n-1)\Bigr]\biggr\},\label{expandH}
\eea
\end{widetext}
where $u_\alpha(m,n)\equiv u_{\langle jl\rangle_\alpha}$ for a nearest-neighbor bond of type $\alpha=x,y,z$ pointing from site $j\in$~A to site $l\in$~B. 
We next define the $2N$-dimensional Majorana vector,
$V=\left(c_{\rm A},c_{\rm B}\right)^T$, with
\be
c_\lambda=\left(c_\lambda(1,1),\ldots c_\lambda(L,1), c_\lambda(1,2),\ldots,
c_\lambda(L,L)\right)^T, \label{vector}
\ee
as well as a complex fermion for each unit cell,
$f(m,n)=\frac12[c_{\rm A}(m,n)-ic_{\rm B}(m,n)]$. With a vector $f$ formed as in Eq.~\eqref{vector}, the transformation between both representations is 
given by 
\be
V=\left(\begin{array}{c}
c_{\rm A}\\ c_{\rm B}
\end{array}\right)=T\left(\begin{array}{c}
f\\ f^\dagger
\end{array}\right),\quad T=\left(\begin{array}{cc}
 {\mathbb 1}_N&{\mathbb 1}_N \\
i {\mathbb 1}_N& -i{\mathbb 1}_N\end{array}\right),
\ee
with  the $N\times N$ identity ${\mathbb 1}_N$. The projection ${\cal P}$ defined in the main text here implies a parity constraint \cite{Pedrocchi2011,Zschocke2015}  for
the total number $N_f$ of $f$ fermions and the total number $N_\chi$ of bond fermions, 
$(-1)^{N_f+N_\chi}=1.$
Here, bond fermion operators are defined as $\chi^{}_{\langle jl\rangle_\alpha}=\frac12\left(c^\alpha_j-ic^\alpha_l\right)$ \cite{Baskaran2007}, such that 
the spin operator $S_j^\alpha =\frac{i}{2}c_j \left(\chi^{\phantom\dagger}_{\langle jl\rangle_\alpha}+\chi^\dagger_{\langle jl\rangle_\alpha}\right)$ 
changes the occupation number of the bond fermion.  In terms of gauge invariant objects, $S_j^\alpha$ flips the two plaquette operators $W_p$ adjacent to this bond.
Using the $f$ fermions, $H$ assumes a Bogoliubov-de-Gennes (BdG) form,
\be\label{BdG}
H=(f^\dagger \; f)\;T^\dagger \left(\begin{array}{cc}\mc H_{\rm AA}&\mc H_{\rm AB}\\
\mc H_{\rm BA}&\mc H_{\rm BB}\end{array}\right) T\left(\begin{array}{c} f\\f^\dagger\end{array}\right),
\ee
where the $N\times N$ matrices $\mc H_{\lambda\lambda'}$ follow from Eq.~(\ref{expandH}).
We next apply a unitary transformation,
$\left(\begin{array}{c} f\\f^\dagger\end{array}\right)=U\left(\begin{array}{c} a\\a^\dagger\end{array}\right)$,
in order to diagonalize Eq.~\eqref{BdG} for a given gauge configuration $|\mc G\rangle$ in terms of new $a$ fermions,
$H=\sum_{\nu=1}^{N}\varepsilon_\nu \left(2 a^\dagger_{\nu}a^{\phantom\dagger}_\nu-1\right),$
where $\varepsilon_\nu$ are the non-negative eigenenergies ordered as $\varepsilon_1<\varepsilon_2<\dots<\varepsilon_N$.
Taking the g.s., $|\mc M_0(\mc G)\rangle$ with $a_\nu|\mc M_0(\mc G)\rangle=0$ for all $\nu$, the g.s.~energy is
$E_0=-\sum_{\nu=1}^{N}\varepsilon_\nu$.
However, one may have to add a fermion to the $\varepsilon_1$ level to fulfill the above parity constraint, resulting in the g.s.~energy  $\tilde{E}_0=E_0+2\varepsilon_1.$
For the two-vortex case, there are two MZMs at zero energy when the vortices are far away, resulting in $\varepsilon_1=0$ and $\tilde{E}_0=E_0$.   
For the uniform zero-vortex state, the g.s.~energy follows by Fourier transformation.  In the thermodynamic limit, one finds  
$\frac{E_0}{N}=-\frac{\sqrt3}{4\pi^2}\int_{\frac12{\rm BZ}}d^2{\mb k}\,\varepsilon(\mb k)$,
where $\frac12$BZ denotes half the Brillouin zone and the dispersion is given by
\be
\varepsilon(\mb k)=\frac{1}{4}\sqrt{\left| K\sum_{i=1}^3e^{i\mb k \cdot \mb a_i}\right|^2+\left(\kappa\sum_{i=1}^3\sin(\mb k\cdot \mb b_i)\right)^2}\label{gszeroflux}
\ee 
with $\mb a_1=\left(0,\frac{1}{\sqrt3}\right)$, $\mb a_2=\left(-\frac12,-\frac1{2\sqrt3}\right)$,   $\mb a_3=\left(\frac12,-\frac1{2\sqrt3}\right)$, and $\mb b_1=\mb a_2-\mb a_3$, $\mb b_2=\mb a_3-\mb a_1$, $\mb b_3=\mb a_1-\mb a_2$ \cite{Kitaev2006}. 
The threshold energy for two-fermion excitations in the zero-vortex sector is given by $\Delta E_{2f}=2\Delta E_{f}=
2\varepsilon(\mb k_0)\propto |\kappa|$, see Eq.~\eqref{d2f} below, 
where $\mb k_0=(2\pi/3,2\pi/\sqrt3)$ is the momentum at the K point of the Brillouin zone. 

We now turn to the spin correlations. Consider first the correlation for $\alpha=z$ within a unit cell (choosing $j\in$~A), where
Eq.~(4) in the main text gives
$C_{\langle jl\rangle_z}=-iu_z(m,n)\langle c_{\rm A}(m,n)c_{\rm B}(m,n)\rangle$.
Labeling the components of the vector \eqref{vector} by the index
$r=r(m,n)=m+(n-1)L$, we obtain $C_{\langle jl\rangle_z}=-iu_z(m,n)\langle V^\dagger W^z_r V\rangle$
with the matrix $W^z_r=\left(\begin{array}{cc}0&Z^{{\rm AB}}_r\\0&0\end{array}\right)$.  
Similarly, we define matrices $W^{\alpha=x,y}_{r}$ by replacing the
diagonal $N\times N$ diagonal matrix $Z^{\rm AB}_r$, 
with only one nonzero matrix element $(Z^{\rm AB}_r)_{r_1r_2}=\delta_{r_1r}\delta_{r_2r}$, 
by 
\bea\nonumber
(X^{\rm AB}_r)_{r_1r_2}&=&\delta_{m_1m}\delta_{n_1n}\delta_{m_2,m+1}\delta_{n_2n},\\
(Y^{\rm AB}_r)_{r_1r_2}&=&\delta_{m_1m}\delta_{n_1n}\delta_{m_2m}\delta_{n_2,n+1}.\label{XAB}
\eea
Using $V^\dagger=\mc A^\dagger U^\dagger T^\dagger$ with $\mc A^\dagger\equiv (a^\dagger,a)$, we obtain
\be
C_{\langle jl\rangle_\alpha}=-iu_\alpha(m,n)\sum_{\nu,\nu'=1}^{2N} (U^\dagger T^\dagger W^\alpha_r TU)_{\nu\nu'}  
\langle \mc A_\nu^\dagger \mc A^{\phantom\dagger}_{\nu'}\rangle.
\ee
For the g.s.~at given $|{\cal G}\rangle$, we thus arrive at
\be
C_{\langle jl\rangle_\alpha}=-iu_\alpha(m,n)\sum_{\nu=N+1}^{2N} (U^\dagger T^\dagger W^\alpha_{r(m,n)} TU)_{\nu\nu} .\label{corrgeneral}
\ee 
If the g.s.~has an occupied $\varepsilon_1$ level because of the parity constraint, we instead find
 \bea\nonumber
\tilde{C}_{\langle jl\rangle_\alpha}&=&-iu_\alpha(m,n)\Bigl[ (U^\dagger T^\dagger W^\alpha_{r(m,n)} TU)_{NN}\\
&&+\sum_{\nu=N+2}^{2N} (U^\dagger T^\dagger W^\alpha_{r(m,n)} TU)_{\nu\nu} \Bigr] .
\eea
In any case, the calculation of spin correlations has been reduced to determining the unitary $U$ diagonalizing the BdG Hamiltonian.

\begin{figure}
\begin{center}
\includegraphics[width=\columnwidth]{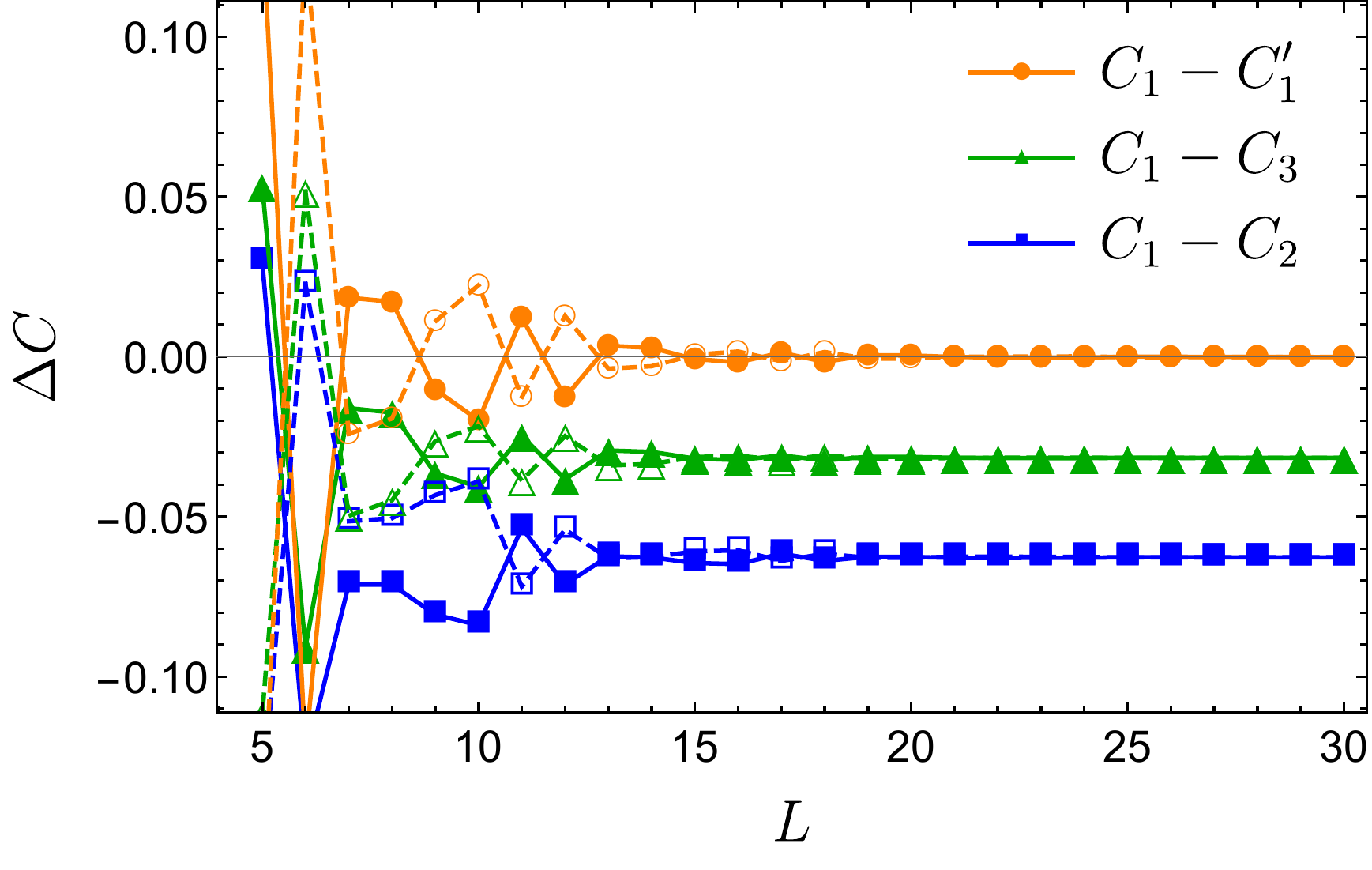}
\end{center}
\caption{Ground state spin correlations $C_1^\prime$ and $C_{1,2,3}$ [see Eq.~(4) and Fig.~1(a) in the main text] vs $L$ (in units of the lattice spacing) for the g.s. with two vortices kept at maximal distance in a system of size $L\times L$.  The parameters are as in Fig.~2(b) of the main text. Filled symbols and dashed lines represent the result for the g.s., $|\mathcal M_0(\mathcal{G})\rangle$, which obeys $a_\nu|\mathcal M_0(\mathcal{G})\rangle=0$ for all $\nu$. Empty symbols and dashed lines represent the result for the state in which the single-particle level with  energy $\varepsilon_1$  is occupied. Note that for both states,
spin-spin correlations converge to the same values in the large-$L$ limit.     }  
\label{fig1SM}
\end{figure}

Figure \ref{fig1SM} shows the differences between spin correlations $C_1^\prime$ and $C_{1,2,3}$ near a vortex as defined in Fig.~1(a) of the main text.  
We study the g.s.~of a system of size $L\times L$ with two vortices kept at maximal distance $d=\lfloor (L-1)/2 \rfloor$.
For the parameters in Fig.~\ref{fig1SM}, the thermodynamic limit (with well separated vortices) is reached for $L\agt 20$.  
The spin correlations $C_1^\prime$ and $C_1$ then become identical, and the charge density profile is isotropic around the vortex center. 
The difference $C_1-C_3$ determines the charge imbalance on sites surrounding a vortex, where $C_1-C_3\simeq -0.0315$ in the thermodynamic limit.

Finally, the calculation of the dynamic response function quoted in Eq.~(11) of the main text involves 
matrix elements between the ground state and excited states. The corresponding spectral function has the form
\be
\mc  A(\omega)=2\pi  \sum_{1\leq\nu<\nu'\leq N}\left|\Lambda_{\nu\nu'} \right|^2  \delta(\omega -\varepsilon_\nu-\varepsilon_{\nu'}),
\ee
where the matrix element is given by \bea
\Lambda_{\nu\nu'}&=&-\frac{i}4\sum_{\ell=1}^6 u_{\alpha_\ell}(m_\ell,n_\ell )\left\{\left[U^\dagger T^\dagger W^{\alpha_\ell}_{r(m_\ell,n_\ell)} TU\right]_{\nu,N+\nu'}\right.\nonumber\\
&&\left.-\left[U^\dagger T^\dagger W^{\alpha_\ell}_{r(m_\ell,n_\ell)} TU\right]_{\nu',N+\nu} \right\}.
\eea
Here $(m_\ell,n_\ell,\alpha_\ell)$ with $\ell=1,\dots,6$ label the unit cells and bond types for the six nearest-neighbor bonds with one site in the hexagon containing the vortex and the other site  outside  the hexagon. If the bond marked by $C_2$ in Fig. 1(a) of the main text corresponds to $(m_0,n_0,x)$, the other five bonds in clockwise order are $(m_0,n_0-1,y)$, $(m_0-1,n_0,z)$, $(m_0-2,n_0+1,x)$, $(m_0-1,n_0+1,y)$, and $(m_0,n_0+1,z)$. The MZM peak for $N_V=4$ in Fig. 4 of the main text occurs at the energy $\varepsilon_{M}=\varepsilon_1+\varepsilon_2$ and its weight  is proportional to $|\Lambda_{12}|^2$. In the four-vortex sector, both eigenenergies  $\varepsilon_1$ and $\varepsilon_2$   decrease exponentially with the inter-vortex distance.

\section{II. Hubbard-Kanamori model and local charge operator}

\subsection{A. Model}
\label{sec2a}

We consider the   Hubbard-Kanamori  model for $d^5$ electrons in an edge-sharing octahedral environment \cite{Jackeli2009}, 
see also Refs.~\cite{Rau2014,Rau2016,Winter2016}.
The five $d$-electrons of the Ru$^{3+}$ ions in a cubic crystal field occupy three $t_{2g}$ orbitals $(xy,yz,zx)$, denoted below by the    complementary index $\alpha=(z,x,y)$, respectively. With the electron creation operator $d^\dagger_{i\alpha\sigma}$ at site $i$ for spin $\sigma$, and using $d^\dagger_{i\alpha}=(d^\dagger_{i\alpha\uparrow},d^\dagger_{i\alpha\downarrow})$,
 we define the electron number operator at this site, $N_i=\sum_\alpha d^\dagger_{i\alpha} d^{\phantom\dagger}_{i\alpha}$,
the spin operator, $\mb S_i=\frac12\sum_{\alpha}d^\dagger_{i\alpha} \boldsymbol\sigma d^{\phantom\dagger}_{i\alpha}$, 
and the orbital angular momentum operator, $\mb  L_i=\sum_{\alpha\beta}d^\dagger_{i\alpha} (\mb l)_{\alpha\beta}d^{\phantom\dagger}_{i\beta}$.
Here $\mb l=(l^x,l^y,l^z)$ represents the $l_{\text{eff}}=1$ orbital angular momentum of the $t_{2g}$ states. In the orbital basis $\{|x\rangle,|y\rangle,|z\rangle\}$, 
 \be
l^x=\left(\begin{array}{ccc}0&0&0\\
0&0&-i\\
0&i&0
\end{array}\right),\  l^y=\left(\begin{array}{ccc}0&0&i\\
0&0&0\\
-i&0&0
\end{array}\right),  \ l^z=\left(\begin{array}{ccc}0&-i&0\\
i&0&0\\
0&0&0
\end{array}\right)\label{lmatrices}
\ee

The Hubbard-Kanamori Hamiltonian \cite{Rau2016,Winter2016}, 
\be H=H_{0}+H_{\rm at}+H_{\rm so},
\ee
contains an orbital- and bond-dependent hopping term, 
\be \label{t2orbital}
H_0 = t\sum_{\langle ij\rangle_\gamma} d^\dagger_{i\alpha} d^{\phantom\dagger}_{j\beta}  
+t'\sum_{\langle\langle ij\rangle\rangle_\gamma} d^\dagger_{i\alpha} d^{\phantom\dagger}_{j\beta} +(\alpha\leftrightarrow \beta),
\ee
where $t$ ($t'$) are the dominant (next-)nearest-neighbor hopping amplitudes, see Fig.~1(b) in the main text. 
These real-valued positive amplitudes refer to transitions between $t_{2g}$ orbitals after integrating out 
the $p$-orbitals at the ligand (Cl) sites. Allowing for a local electrostatic potential shift $V_0$ induced by a probe tip voltage,
the atomic on-site Hamiltonian, $H_{\rm at}=\sum_i H_{{\rm at}}^{(i)}$, and the local
spin-orbit term, $H_{\rm so}=\sum_i H_{\rm so}^{(i)}$,  are respectively given by  
\bea\nonumber
H_{{\rm at}}^{(i)} &=&\frac{U-3J_H}2(N_i-5)^2-2J_H\mb S_i^2-\frac{J_H}2\mb L_i^2 -
eV_0 N_i, \\
H_{\rm so}^{(i)}&=& -\lambda_{\rm so} \mb S_i\cdot \mb L_i,\label{atomicH}
\eea
where $U$ is the on-site Coulomb repulsion, $0<J_H<U/3$ denotes the Hund coupling, and $\lambda_{\rm so}>0$ is the spin-orbit coupling.

We consider the strong-coupling regime with $t,t' \ll U,J_H$ throughout. For  $\lambda_{\rm so}=0$,   the g.s.~of $H_{\rm at}^{(i)}$ is six-fold degenerate and has 
the quantum numbers $N_i=5$, $S_i=1/2$, and $L_i=1$.  
In order to simplify the calculations below, we next perform a particle-hole transformation,
\be\label{phtrafo}
d_{i\alpha}=\left(\begin{array}{c}d_{i\alpha\uparrow}\\ d_{i\alpha\downarrow}
\end{array}\right) = \left(\begin{array}{c}h^\dagger_{i\alpha\downarrow}\\
-h^\dagger_{i\alpha\uparrow}
\end{array}\right)=i\sigma^y\left(h^\dagger_{i\alpha}\right)^T.
\ee
In terms of the hole operators $h_{i\alpha}$, we then have the on-site operators
\bea\label{onsiteop1}
\bar N_i&\equiv& 6-N_i =\sum_\alpha h^\dagger_{i\alpha}h^{\phantom\dagger}_{i\alpha},\\ \nonumber
\mb S_i&=&\frac12\sum_{\alpha}h^\dagger_{i\alpha} \boldsymbol\sigma h^{\phantom\dagger}_{i\alpha}, \quad
\mb L_i=-\sum_{\alpha\beta}h^\dagger_{i\alpha} (\mb l)_{\alpha\beta}h^{\phantom\dagger}_{i\beta}.
\eea
At low energies, $H_{\rm at}$ in Eq.~\eqref{atomicH} implies that we have $\bar N_i=1$ hole per site. Note that $H_0$ in Eq.~\eqref{t2orbital} effectively changes sign after the particle-hole transformation.

Below it is convenient to use the index $s=(\alpha,\sigma)=1,\ldots,6$, and combine the orbital and spin degrees of freedom in a 
six-component spinor for each site $i$:
\be\label{s6}
h_i^\dagger=\left (h^{\dagger}_{ix\uparrow},h^{\dagger}_{iy\uparrow},h^{\dagger}_{iz\uparrow},h^{\dagger}_{ix\downarrow},h^{\dagger}_{iy\downarrow},h^{\dagger}_{iz\downarrow}\right).
\ee 
The hopping amplitudes in $H_0$ are thereby expressed in terms of $6\times 6$ matrices $\mb T_{ij}$ with the matrix elements
\be\label{tij}
\left(\mb T_{ij}\right)_{ss'}=\Bigl(\mathbb{1}_2\otimes \mb T^{(o)}_{ij}\Bigr)_{ss'}
\ee
where $\mathbb{1}_2$ is the identity in spin space and the matrix  $\mb T_{ij}^{(o)}$ 
in orbital space depends on the bond type of the link between $i$ and $j$.  Specifically, for nearest-neighbor bonds, 
\bea \nonumber
\mb T_{\langle ij\rangle_x}^{(o)}&=&t\left(\begin{array}{ccc}
0&0&0\\
0&0&1\\
0&1&0
\end{array}\right),\quad
\mb T_{\langle ij\rangle_y}^{(o)}=t\left(\begin{array}{ccc}
0&0&1\\
0&0&0\\
1&0&0
\end{array}\right),\\
\mb T_{\langle ij\rangle_z}^{(o)}&=&t\left(\begin{array}{ccc}
0&1&0\\
1&0&0\\
0&0&0
\end{array}\right).\label{Tmatrices}
\eea
For next-nearest neighbors, the $\mb T^{(o)}_{ij}$ matrices follow 
from Eq.~\eqref{Tmatrices} by replacing $t\to t'$.  Moreover, from Eq.~\eqref{onsiteop1}, one obtains with $\mb l$ in Eq.~\eqref{lmatrices}:
\be
\bar N_i=h^\dagger_{i}h^{\phantom\dagger}_i,\quad \mb S_i=\frac12h^\dagger_{i}(\boldsymbol \sigma\otimes \mathbb{1}_3)h^{\phantom\dagger}_i,
\quad \mb L_i=-h^\dagger_{i}( \mathbb{1}_2\otimes \mb l)h^{\phantom\dagger}_i.
\ee

The spin-orbit coupling, $H_{\rm so}$, then partially lifts the  degeneracy of the atomic Hamiltonian $H_{\rm at}$ by splitting the $t_{2g}$ states into two multiplets of total angular momentum $j_{\text{eff}}=1/2$ and  $j_{\text{eff}}=3/2$,  respectively.
 Since the particle-hole transformation \eqref{phtrafo} effectively reverses the sign of the spin-orbit term in Eq.~(\ref{atomicH}), the g.s.~corresponds to a hole in the $j_{\text{eff}}=1/2$ doublet. We here follow  Refs.~\cite{Jackeli2009,Rau2014} and implement the projection onto the low-energy subspace in two steps. First, we derive  the effective operators in the strong-coupling  regime obtained by considering  the subspace with one hole in a $t_{2g}$ orbital and neglecting the spin-orbit coupling. Second, we include the spin-orbit coupling by taking the matrix elements of  the effective operators between states in the $j_{\text{eff}}=1/2$ subspace, which is spanned by
\bea
 |+\rangle&=&\frac1{\sqrt{3}}\left(-|z,\uparrow\rangle-i|y,\downarrow\rangle-|x,\downarrow\rangle\right),\nonumber \\
|-\rangle&=&\frac1{\sqrt{3}}\left(|z,\downarrow\rangle+i|y,\uparrow\rangle-|x,\uparrow\rangle\right).
\eea
This approach is valid provided that the energy scales in $H$ satisfy the condition \cite{Jackeli2009,Rau2016}
\be \label{scalesep}
 t,t'\ll \lambda_{\text{so}}\ll U,J_H.
\ee 
 
With the projector $\mc P^{(n)}_i$ onto the subspace with $n$ holes at site $i$, the low-energy projection to a single hole per site is implemented by 
\be\label{plow}
\mc P_{\rm low}=\prod_i \mc P_i^{(1)}.
\ee
Since we consider processes up to third order in $(t,t')/U$ below, the g.s.~never couples to states with more than
two holes per site.  We can thus approximate the identity at site $i$ by $\mathbb{1}_i \simeq \mc P_i^{(0)}+ \mc P_i^{(1)}+ \mc P_i^{(2)}$, and write the hopping term as
\be
H_0=T_0+T_1+T_{-1},
\ee
with
 \bea
T_0&=&-\sum_{  ij}\left[\mc P_i^{(1)}h^\dagger_{i}  {\mb T}_{ij}h^{\phantom\dagger}_{j}  \mc P_j^{(1)}+ \mc P_{i}^{(2)} h^\dagger_{i} {\mb T}_{ij} h^{\phantom\dagger}_{j}  \mc P_{j}^{(2)}\right],\nonumber \\
T_1&=&-\sum_{  ij}  \mc P_i^{(2)}h^\dagger_{i} {\mb T}_{ij}h^{\phantom\dagger}_{j}  \mc P_j^{(1)}  ,\quad T_{-1}=T_1^\dagger,
\eea
where we used the relations $h_i^{} \mc P_i^{(0)}=\mc P_i^{(0)} h_i^\dagger=0$.

Following Refs.~\cite{Takahashi1977,MacDonald1988,Chernyshev2004}, our goal will be to identify a canonical transformation,
\be \label{canonical}
\tilde H = e^S H e^{-S} = H+[S,H]+\frac12 [S,[S,H]]+\cdots,
\ee
such that all terms leaving the low-energy space with one hole per site are eliminated from the low-energy Hamiltonian up to the desired order in perturbation theory.
This procedure is equivalent to performing a Schrieffer-Wolff transformation.  Once the generator $S$ has been determined,
an arbitrary local operator ${\cal O}_i$ commuting with $\bar N_i$ is represented by the transformed operator
\bea\nonumber
\tilde {\cal O}_i &=& e^S {\cal O}_i e^{-S} = {\cal O}_i+[S,{\cal O}_i]+\frac{1}{2!} [S,[S,{\cal O}_i]]+\\&+&\frac{1}{3!} 
[S,[S,[S,{\cal O}_i]]]+\cdots.\label{operatortransform}
 \eea
By setting ${\cal O}_i=h_i^\dagger h_i^{}-1$, Eq.~\eqref{operatortransform} determines the low-energy form of the charge imbalance operator, see Sec.~\ref{sec2b}.

To proceed, we split the projection operators $\mc P^{(n)}_i$ with $n=0,1,2$ into channels with different orbital angular momentum, 
\be 
\mc P^{(n)}_{i}=\sum_{L=0}^2\mc P^{(n)}_{i,L}.
\ee
Writing $H_{\rm at}=V_\rho+V_\sigma+V_\ell$ in Eq.~\eqref{atomicH}, with operator contributions due to density ($\rho$), spin ($\sigma$), and orbital angular momentum ($\ell$) terms,
we then compute the eigenvalue changes, $\Delta E=\Delta V_\rho+\Delta V_\sigma+\Delta V_\ell$,
from the basic commutator relation ($n,n'=0,1,2$)
\bea \nonumber
&& \sum_{ij}\left[H_{\rm at},\mc P^{(n)}_{i,L}h_{i}^{ \dagger}{\mb T}_{ij}h_j\mc P^{(n')}_{j,L'}\right]= \\ 
&& \sum_{ij}\Delta E(n,L;n',L')\,\mc P^{(n)}_{i,L}h_{i}^{ \dagger}{\mb T}_{ij}h_j\mc P^{(n')}_{j,L'},\label{simplerule}
\eea
where the orbital quantum numbers $L,L'$ have to be compatible with $n,n'$.
The result for $V_\rho$ is independent of $L$ and $L'$, whereas the result for $V_\sigma$ depends only on the parity of $L$ and $L'$ since the two holes form a singlet (triplet) for even (odd) $L$. For $n=1$, only $L=1$ is allowed and we can then omit the index $L$.  
In particular, one finds the relation 
\be\label{QQ}
\sum_{ij}\left[H_{\rm at},\mc P^{(2)}_{i,L}h_{i}^{ \dagger}{\mb T}_{ij}h_j\mc P^{(1)}_{j}\right]=
\sum_{ij}\Delta E_L\,\mc P^{(2)}_{i,L}h_{i}^{ \dagger}{\mb T}_{ij}h_j\mc P^{(1)}_{j},
\ee
where the energy shifts $\Delta E_L\equiv \Delta E(n=2,L;n'=1,L'=1)$ are for $V_0=0$ given by
\be \label{delL}
\Delta E_0=U+2J_H,\quad \Delta E_1=U-3J_H, \quad \Delta E_2=U-J_H.
\ee
For a potential difference $V_0\ne 0$ between sites $i$ and $j$,
one has to replace $\Delta E_L\to \Delta E_L \pm V_0$ for the $(ij)$ and $(ji)$ terms in Eq.~\eqref{QQ}, respectively.

The generator of the canonical transformation to first order in $(t,t')/U$ takes the form $S_1=S_1^{(+)}-S_1^{(-)}$, with $S_1^{(-)}=\left[S_1^{(+)}\right]^\dagger$ and
\be
S_1^{(+)}=-\sum_{L=0}^2 \frac{1}{\Delta E_L} \sum_{ij} \mc P^{(2)}_{i,L}h_{i}^{ \dagger}{\mb T}_{ij}h_j\mc P^{(1)}_j.\label{S1plus}
\ee 
 We then find
\be\label{cond1} 
[H_{\rm at},S_1]=T_1+T_{-1}.
\ee 
Under the condition \eqref{cond1}, all  first-order terms are eliminated from the transformed Hamiltonian in Eq.~\eqref{canonical}.  

Using Eq.~\eqref{S1plus} and taking into account the spin-orbit coupling by projecting the transformed Hamiltonian to the $j_{\rm eff}=1/2$ sector, 
the effective spin Hamiltonian at second order in $(t,t')/U$
is then given by the Kitaev honeycomb model with a ferromagnetic exchange coupling $K\propto J_H$ as specified in the main text, plus next-nearest Kitaev couplings $\propto (t')^2/U$.
Importantly, the standard Heisenberg interaction is absent. We thus recover the seminal results of Ref.~\cite{Jackeli2009}.

Specifically, for $V_0=0$, the above calculation reproduces Eq.~(4) of Ref.~\cite{Rau2014} upon setting $t_2=t$ and $t_1=t_3=0$ in their equations. 
It is straightforward to include their $t_1$ and $t_3$ couplings in the $\mb T_{ij}$ matrices in Eq.~\eqref{tij}. However, such terms break the integrability of the Kitaev model,
see Sec.~\ref{sec2b4} below.  Let us also note that the inclusion of magnetic field effects is discussed in Sec.~\ref{sec2b3} below.
Finally, for $V_0\ne 0$, we find by similar steps as outlined above that the exchange coupling $K$ is replaced by $K(V_0)$ in Eq.~(9) of the main text.

\subsection{B. Local charge imbalance operator}\label{sec2b}

Let us now turn to the derivation of the low-energy form of the local charge imbalance operator, $e\delta \hat n_i$, at site $i$.  We start from Eq.~\eqref{operatortransform}
with ${\cal O}_i= h_i^\dagger h_i^{}-1$ and use the low-energy projection operator ${\cal P}_{\rm low}$ in Eq.~\eqref{plow}.  The operator $\delta \hat n_i$ then follows as
\be\label{dni}
\delta \hat n_i={\cal P}_{\rm low}\tilde {\cal O}_i {\cal P}_{\rm low}.
\ee
Using the generator $S=S_1+S_2$ of the canonical transformation, with $S_1$ in Eq.~\eqref{S1plus} and $S_2$ in Eq.~\eqref{S2plus} below, and systematically retaining all terms
 up to third order in $(t,t')/U$, we find
\bea \label{lowdni}
\delta \hat n_i &=& \frac12{\cal P}_{\rm low}\left[ S_1, [S_1,{\cal O}_i]\right] {\cal P}_{\rm low}+ \\ \nonumber 
&+& \frac12 {\cal P}_{\rm low}\left[ S_1, [S_2,{\cal O}_i]\right]{\cal P}_{\rm low}+\frac12{\cal P}_{\rm low}\left[ S_2, [S_1,{\cal O}_i]\right] {\cal P}_{\rm low},
\eea
where we have exploited the relations
\be\label{symm}
{\cal P}_{\rm low} {\cal O}_i {\cal P}_{\rm low} = {\cal P}_{\rm low} \left[ S,{\cal O}_i \right] {\cal P}_{\rm low} = 0.
\ee
Importantly, Eq.~\eqref{symm} implies that the third-order expression for $\delta\hat n_i$ is independent of the generator $S_3$, see Eq.~\eqref{lowdni}.

For vanishing local electrostatic potential, i.e., for $V_0=0$, we find that the second-order term, i.e., the first term on the r.h.s.~of Eq.~\eqref{lowdni}, vanishes.
However, for $V_0\ne 0$, it produces a finite contribution to $\delta\hat n_i$ since the potential difference between sites $i$ and $j$ breaks the symmetry between a pair of sites on a given bond. 
The second-order term then dominates over the remaining third-order contributions in Eq.~\eqref{lowdni}. 

\subsubsection{B.1. Finite local tip voltage}\label{sec2b1}

We start with the case of a finite electrostatic potential $V_0\ne 0$ at site $i$, which could, for instance, be generated by means of a voltage-biased STM tip.
The third-order contributions in Eq.~\eqref{lowdni} can then be neglected and one arrives at
\be \label{ni2}
\delta \hat n_i = \frac12{\cal P}_{\rm low}\left[ S_1, [S_1,{\cal O}_i]\right]{\cal P}_{\rm low}.
\ee
\\
Using $S_1$ in Eq.~\eqref{S1plus}, with the energy shifts  $\Delta E_L\to \Delta E_L\pm V_0$ discussed above, Eq.~\eqref{ni2} can be written as a sum over the bonds $\langle ij\rangle_\alpha$,
i.e., $\delta\hat n_i=\sum_j \delta \hat n_{i,j}^{(2)}$, where the pair-wise contributions arise at second order in $(t,t')/U$: 
\bea \nonumber
\delta \hat n_{i,j}^{(2)} &=& \sum_{L=0}^2\biggl [\frac{1}{(\Delta E_L+eV_0)^2}
h_j^\dagger {\mb T}_{ji}h_i^{\phantom\dagger}\mc P_{i,L}^{(2)} h_i^\dagger{\mb T}_{ij}h_j^{\phantom\dagger}
\\ &-& \frac{1}{(\Delta E_L-eV_0)^2}h_i^\dagger{\mb T}_{ij}h_j^{\phantom\dagger}\mc P_{j,L}^{(2)} h_j^\dagger{\mb T}_{ji}h_i^{\phantom\dagger}\biggr].\label{oppp}
\eea
Note that $\delta \hat n_{i,j}^{(2)}=-\delta \hat n_{j,i}^{(2)}$.  The superscript indicates that we have only pair-wise contributions resulting from the second-order expansion.

Let us specify the matrix elements of the operator in Eq.~\eqref{oppp} in the basis $|s_i,s_j\rangle$, with the index $s=(\alpha,\sigma)= 1,\ldots,6$ and the spinor operator components $h_{i,s}$, see Eq.~\eqref{s6}.
We find
\begin{widetext}
\be\label{matdelt}
\langle s'_i,s_j' |\delta\hat n^{(2)}_{i,j} |s_i,s_j\rangle=\sum_{s_2,s_3=1}^6\sum_{L=0}^2\left[\frac{({\mb T}_{ji})_{s_j's_2}({\mb T}_{ij})_{s_3s_j}}{(\Delta E_L+eV_0)^2} F_L(s_i',s_2,s_3,s_i)
-\frac{({\mb T}_{ij})_{s_i's_2}({\mb T}_{ji})_{s_3s_i}}{(\Delta E_L-eV_0)^2} F_L(s_j',s_2,s_3,s_j)\right],
\ee
where we have used $\langle s_i^\prime | h_{i,s}^\dagger h_{i,s'}^{}|s_i\rangle=\delta_{s_i^\prime,s} \delta_{s_i,s'}$ and the functions 
$F_L(s_1,s_2,s_3,s_4)=\langle s_1| h^{}_{i,s_2} {\cal P}_{i,L}^{(2)} h^\dagger_{i,s_3} |s_4\rangle.$   Using the notation $\bar\sigma=-\sigma$,
their explicit form is given by
\bea
F_0(s_1,s_2,s_3,s_4)&=&\frac{\sigma_2\sigma_3}{3}  \delta_{\alpha_1 \alpha_2}\delta_{\alpha_3\alpha_4} \delta_{\sigma_2\bar\sigma_1}\delta_{\sigma_3\bar\sigma_4},\nonumber\\
F_1(s_1,s_2,s_3,s_4)&=&\frac12(\delta_{\alpha_2\alpha_3}\delta_{\alpha_1\alpha_4}-\delta_{\alpha_2\alpha_4}\delta_{\alpha_1\alpha_3})(\delta_{\sigma_2\sigma_3}\delta_{\sigma_1\sigma_4}+\delta_{\sigma_2\sigma_4}\delta_{\sigma_3\sigma_1}),\nonumber\\ \label{FLfunc}
F_2(s_1,s_2,s_3,s_4)&=&  \delta_{s_2s_3}\delta_{s_1s_4}-\delta_{s_1s_3}\delta_{s_2s_4}-F_0(s_1,s_2,s_3,s_4)-F_1(s_1,s_2,s_3,s_4).
\eea
Let us also recall that $\left({\mb T}_{jk}\right)_{s_1s_2}=\left(  {\mb T}^{(o)}_{jk}\right)_{\alpha_1 \alpha_2}\delta_{\sigma_1\sigma_2}$, see Eq.~\eqref{tij}.

In the final step, we project the matrix representation for $\delta\hat n_i$ in Eq.~\eqref{matdelt} to the $j_{\rm eff}=1/2$ subspace in order to take into account the spin-orbit coupling.
We thereby arrive at Eq.~(10) in the main text, with the dimensionless functions
\bea
f_0(\xi_0,\eta)&=&\frac{6+71 \eta^4-149 \eta^3+111 \eta^2-39 \eta +(3 \eta -1)^3 (13 \eta -6) \xi_0^4-2 (1-3 \eta )^2 \left(11 \eta^2-17 \eta +6\right) \xi_0^2}{9 (1-3 \eta )^3 (1-\xi_0^2)^2    [(1-\eta)^2 -(1-3 \eta )^2 \xi_0^2]^2}, \nonumber\\
f_s(\xi_0,\eta)&=&\frac{4\eta  \left[3-13 \eta^3+25 \eta^2-15 \eta +(3 \eta -1)^3 \xi_0^4+2 (\eta -1) (1-3 \eta )^2 \xi_0^2\right]}{3 (1-3 \eta )^3 (1-\xi_0^2)^2  [(1-\eta)^2 -(3 \eta -1)^2 \xi_0^2]^2}.\label{gfuncion}
\eea
Here $\xi_0=eV_0/[(1-3\eta)U]$ is the dimensionless voltage parameter, and the dimensionless Hund coupling is $\eta=J_H/U$ with $0<\eta<1/3$.  Note that $\delta \hat n^{(2)}_i=0$ for $V_0=0$, 
see Eq.~(10) in the main text.

By similar steps, one finds that the function $w(\xi_0,\eta)$ in Eq.~(11) of the main text is given by 
\be
w(\xi_0,\eta)=\frac{(1-\eta)^2 ( 3 -12\eta+ 13 \eta^2)-2 \left(1-4 \eta+3 \eta^2 \right)^2\xi_0^2- (1-\eta) (1-3 \eta )^3\xi_0^4}{\left(1-\xi_0^2\right)^2 \left[(1- \eta)^2 -(1-6 \eta+9 \eta^2 )\xi_0^2\right]^2},
\ee
\end{widetext}

The local voltage $V_0$ may be applied via an STM tip. One can then detect whether a $\mathbb{Z}_2$ vortex has been trapped 
underneath the tip by performing the energy
absorption spectroscopy illustrated in Figs.~3 and 4 in the main text.  Apart from the Majorana peak at very low energy, a 
clear signature is that, once the vortex has been trapped, the continuum in the 
absorption spectrum starts at a lower energy (approximately half) than in the 
absence of the vortex. This effect is due to the MZM associated with a $\mathbb{Z}_2$ vortex.
We note that it is not possible to directly measure the  energy difference $\Delta E_{4v}$  (see Fig.~3 of the main text)  since the effective 
charge operator is local and cannot excite four vortices at well separated tip positions.

\subsubsection{B.2. Local charge imbalance operator for $V_0=0$ }\label{sec2b2}

Next we turn to the case $V_0=0$, where no external tip potential is present and therefore no contributions to $\delta \hat n_i$ appear up to the second order in $(t,t')/U$.  
However, at the third order, intrinsic charge imbalance contributions are found from Eq.~\eqref{lowdni}.  We then need the second-order generator $S_2$ 
which is determined from the condition 
\be
[S_1,T_0]+[S_2,H_{\rm at}]=0.
\ee
The solution can be written as $S_2=S_2^{(+)}-S_2^{(-)}$ with $S_2^{(-)}=\left[S_2^{(+)}\right]^\dagger$ and
\begin{widetext}
\be
S_2^{(+)}=\sum_{ijkl}\sum_{n=1,2}\sum_{L,L',L''}\frac{1} {\Delta E_L[\Delta E_L+\Delta E(n,L';n,L'')]} 
\left [ \mc P_{i,L}^{(2)}h^\dagger_{i} {\mb T}_{ij}h^{\phantom\dagger}_{j}\mc P_j^{(1)} , \mc P_{k,L'}^{(n)}h^\dagger_{k} 
{\mb T}_{kl} h^{\phantom\dagger}_{l}\mc P_{l,L''}^{(n)}\right ].\label{S2plus}
\ee
Using $S=S_1+S_2$ in Eq.~\eqref{lowdni} for $V_0=0$, we obtain
\bea
\delta \hat n_l&=&-\sum_{j}\sum_{i'j'k'l'}\sum_{n=1,2}\sum_{L,L',L'',L'''} \frac1{\Delta E_L\Delta E_{L'}[\Delta E_{L'}+\Delta E(n,L'';n,L''')]}  \nonumber\\
&&\times  h^\dagger_{i'}{\mb T}_{i'j'}h^{\phantom\dagger}_{j'}\mc P_{j',L'}^{(2)}\mc P_{k',L'''}^{(n)}h^\dagger_{k'}{\mb T}_{k'l'}h^{\phantom\dagger}_{l'}\mc P_{l',L''}^{(n)}\left(\mc P^{(2)}_{l,L}h_{l}^{ \dagger}{\mb T}_{lj}h_j -\mc P^{(2)}_{j,L}h_{j}^{ \dagger}{\mb T}_{jl}h_l \right)+\text{h.c.},\label{manysums}
\eea
where we have dropped the projection operators on the left and right sides since they equal the identity on the low-energy subspace with one hole per site.
Taking the matrix elements with respect to the three-site basis $|s_j,s_k,s_l\rangle$, see Eq.~\eqref{s6}, we find
\bea
\langle s'_j,s'_k,s'_l |\delta n_l|s_j,s_k,s_l\rangle&=& - \sum_{s_2,s_3,s_4,s_5} ({\mb T}_{jk})_{s_j's_2}( {\mb T}_{kl})_{s_3s_4}({\mb T}_{lj})_{s_5s_j}  \sum_{LL'}\frac{F_{L'}(s_k',s_2,s_3,s_k)F_L(s_l',s_4,s_5,s_l)}{(\Delta E_L)^2\Delta E_{L'}}\nonumber\\
&&  -\sum_{s_2,s_5} ( {\mb T}_{jk})_{s_k's_2}( {\mb T}_{kl})_{s_l's_k}({\mb T}_{lj})_{s_5s_l}\sum_L\frac{F_L(s_j',s_2,s_5,s_j)  } {(\Delta E_L)^3}\nonumber\\
&& +\sum_{s_2,s_3,s_4,s_5}( {\mb T}_{lk})_{s_l's_2}( {\mb T}_{kj})_{s_3s_4}({\mb T}_{jl})_{s_5s_l}\sum_{LL'}  \frac{F_{L}(s_j',s_4,s_5,s_j)F_{L'}(s_k',s_2,s_3,s_k)  }{(\Delta E_L)^2\Delta E_{L'}}  \nonumber\\
&&+\sum_{s_2,s_5}( {\mb T}_{lk})_{s_k's_2}( {\mb T}_{kj})_{s_j's_k}({\mb T}_{jl})_{s_5s_j} \sum_{L}    
\frac{F_L(s_l',s_2,s_5,s_l) }{(\Delta E_L)^3} +(j\leftrightarrow k)  +\text{h.c.},\label{sumdeltanlPL}
\eea
\end{widetext}
with the $F_L$ functions in Eq.~\eqref{FLfunc}.   After projecting these matrix elements to the $j_{\rm eff}=1/2$ sector
favored by the spin-orbit coupling, we finally arrive at Eqs.~(5) and (6) in the main text. 

Concerning the interpretation of these results, it is instructive to compare the local charge imbalance operator in Eq.~(5) in the main text to
the corresponding simpler result for the Hubbard model, see Eq.~(1) in the main text. In the latter case, the effective charge operator 
generated by third-order perturbation theory is completely determined by symmetries. The most important constraint comes from SU$(2)$ 
symmetry which imposes that only scalar products of spin operators can appear. 
In addition, the condition of vanishing charge imbalance for uniform spin-spin correlations fixes the relative coefficients of the three terms in Eq.~(1)
in the main text \cite{Bulaevskii2008,Khomskii2010}.
Our result for the effective charge operator in Kitaev materials obeys only the latter
constraint but is not restricted by SU$(2)$ symmetry. In fact, it is remarkable that only
diagonal two-spin operators appear in Eq.~(5) of the main text even though off-diagonal operators are also
allowed by symmetry. This result is reminiscent of the derivation of the pure Kitaev 
model by Jackeli and Khaliullin \cite{Jackeli2009}.

Similarly, one can understand the sign of the charge imbalance $\rho_j$ on the hexagon sites surrounding a $\mathbb{Z}_2$ vortex. 
To that end, consider a single triangle $(jkl)$ of sites. Both Eqs.~(1) and (5) in the main text predict a negative charge imbalance contribution from this triangle, $\rho_j<0$, if (i) the hopping parameters are positive (i.e., they have the usual sign as compared to the Hubbard model), (ii) the spin-spin correlations are also positive as expected for 
ferromagnetic interactions, and (iii) the spin-spin correlation on the $(kl)$ bond opposite to site $j$ is stronger than on the other two bonds. 
As a consequence, charge carriers tend to move towards stronger exchange bonds. In our case, the hoppings are positive when considering 
the Hamiltonian for holes. Since the spin-spin correlations for the $C_3$ bonds in Fig.~1 of the main text  are stronger than those for
$C_1$ bonds, see Fig.~\ref{fig1SM} above, we conclude that holes tend to move away from the vortex.  We can thereby understand that $\rho_j<0$
for sites adjacent to a vortex.

\subsubsection{B.3. Magnetic field effects}\label{sec2b3}

In our derivation of the local charge imbalance operator, we have assumed that the relevant energy scales in the multi-orbital Hubbard model
show the clear hierarchy in Eq.~\eqref{scalesep}.   In the  above discussion, we have tacitly neglected the effects of an external magnetic field ${\bf h}$
when performing the canonical transformation. 
In order to justify this step, let us first note that the Zeeman energy is typically parametrically small compared to 
the atomic energy scales $(\lambda_{\rm so},J_H,U)$.
Therefore the coupling to ${\bf h}$ does not change the number of electrons at each site,  and
including ${\bf h}$ from the outset would only cause small quantitative corrections to
the energy shifts $\Delta E_L$ in Eq.~\eqref{delL}.  As a consequence, the Zeeman term, $H_Z=- \sum_j h_j S_j$, can be added 
after the derivation of the effective spin model from the multi-orbital Hubbard model. Similarly, 
the local charge imbalance operator is then determined by ${\bf h}$-independent expressions, see  
Eq.~(5) [Eq.~(10)] in the main text for $V_0=0$ [for $V_0\ne 0$].

To study the gapped Kitaev spin liquid in a magnetic field, one may consider 
a magnetic field ${\bf h}$ along the $[111]$ direction, see, e.g., Ref.~\cite{Gordon2019}.
However, the Zeeman term $H_Z$ breaks the integrability of the low-energy model because it does not commute with the plaquette operators $W_p$. 
In order to allow for analytical progress, we follow Kitaev \cite{Kitaev2006} who showed that the 
main effect of time reversal symmetry breaking is to generate a nontrivial mass in the
spectrum of the Majorana fermions.  This effect can be captured by replacing the Zeeman
term by the effective coupling $\propto \kappa$ in Eq.~\eqref{expandH}, where $\kappa \propto h_x h_y h_z/K^2$ follows from
third-order perturbation theory.  More generally,  $\kappa$ represents the leading time reversal symmetry breaking interaction 
which still preserves integrability.  

Let us note that the described charge redistribution effects around a $\mathbb{Z}_2$ vortex are also expected
in the gapless zero-field case with $\kappa\to 0$.  However, the regime of small $\kappa$
is technically more demanding because of strong finite size effects.  For $0.05K\le \kappa \le 0.5K$, 
we have explicitly checked that the spin-spin correlations will numerically converge to their respective thermodynamic limit value
already for relatively small finite-size lattices, see Fig.~\ref{fig1SM} above.  In fact, for those values of $\kappa$, we found no qualitative changes for 
the results reported here.

It is also worth mentioning that $\kappa$ is directly related to the gap for Majorana fermion excitations. Indeed, from Eq.~\eqref{gszeroflux}, 
the two-fermion gap is given by 
\be \label{d2f}
\Delta E_{2f} = \frac{3\sqrt{3} }{4} |\kappa|.
\ee 
Experimental results for $\alpha$-RuCl$_3$ have reported $\Delta E_{2f}\approx K$ for magnetic fields of order $10$~Tesla \cite{Sears2017,Tanaka2020}. 
Equation \eqref{d2f} predicts $\kappa \approx 0.77K$ for $\Delta E_{2f}=K$, which is somewhat larger but overall 
consistent with the values of $\kappa$ studied by us.

In principle, the time reversal symmetry breaking associated with an external magnetic field can also allow for
three-spin terms in the local charge imbalance operator. This effect is already present for the simpler case of the Hubbard model and then 
modifies Eq.~(1) in the main text. 
However, such three-spin terms turn out to be extremely small for realistic fields, where the magnetic flux through a triangular plaquette is 
much smaller than the flux quantum.  We have therefore neglected three-spin contributions to the charge imbalance operator throughout.

\subsubsection{B.4. Towards more microscopic models}\label{sec2b4}

The above projection of the multi-orbital Hubbard-Kanamori model to the integrable Kitaev model involves a number of assumptions. 
We here  discuss  several aspects which may complicate the analysis of experimental data on the charge
redistribution in real Kitaev materials.  Nonetheless, as long as the additional terms in the projected
Hamiltonian (which are neglected in the Kitaev model) remain small, we argue below that the results presented in our manuscript are robust and
will capture the characteristic charge redistribution and absorption spectrocopy features in Kitaev materials arising from the presence of vortices.  

As concrete example, we consider the case of  $\alpha$-RuCl$_3$, where the local moments whose spin configuration can be manipulated 
at low energies are associated with the Ru ions. Their electronic configuration is
described by a single hole with $j_{\rm eff} = 1/2$.   
In general, the projection of the microscopic model to the low-energy spin model will not only give Kitaev interactions
as specified in Eq.~(2) in the main text.  In fact, a more general model will include conventional isotropic Heisenberg couplings as well as the
so-called $\Gamma$ interactions \cite{Rau2014,Rau2016,Winter2016}. 
The local charge imbalance operator may then also pick up additional contributions beyond Eq.~(5) in the main text. 
Some of these subleading contributions can easily be taken into account within the formalism in Sec.~\ref{sec2a}
by allowing for $t_1\ne 0$ and $t_3\ne 0$ (using the notation of Ref.~\cite{Rau2014})  in the hopping matrices~\eqref{Tmatrices}.
However, a more serious obstacle is that these additional interactions will spoil the exact solvability of the Kitaev model.  
A consistent quantitative treatment of  spin-spin correlations and the corresponding charge redistribution effects thus requires further 
analytical approximations and/or the implementation of more powerful numerical methods.  We leave such questions 
to future work.   As long as all additional interactions beyond the Kitaev model remain small, however,   
the results presented here are expected to capture the essential physics.

We next remark that additional terms in the low-energy spin model may also give rise to longer-range spin-spin
correlations beyond Eq.~(4) in the main text, i.e., beyond nearest-neighbor terms. 
Importantly, such contributions are expected to decay exponentially with distance since we are in a gapped phase.
Since the resulting modifications of the spin-spin correlations are exponentially small,  we do not expect qualitative differences
to the shown results, at least if the 
additional terms in $H$ are small.  Moreover, the charge redistribution should be most easily 
detectable in the immediate vicinity of the vortex, where nearest-neighbor 
correlations dominate.

Furthermore, one may argue that the Cl atoms (which are somewhat closer to the STM tip than the Ru atoms) could play an important 
role for the charge redistribution effects described in our work, and may even obscure the signal due to $\mathbb{Z}_2$ vortices 
obtained from the low-energy spin sector.  In order to address this point, let us recall that the Cl states correspond to high-energy electronic levels, 
which are integrated out when projecting the microscopic multi-orbital model to the low-energy spin model \cite{Rau2016}. 
The charge polarizability associated with the Cl atoms is therefore very small.  
The corresponding charge density contribution is expected to show almost no sensitivity to the 
presence or absence of a $\mathbb{Z}_2$ vortex.  

A more serious concern comes from the fact that much of the phenomenology of a Kitaev spin liquid 
may be reproduced from alternative topologically trivial (or at least different) states, e.g., by assuming a trivial partially polarized state.
In fact, even the sign structure of the thermal Hall conductivity, which has been interpreted as a
signature of the gapped Kitaev spin liquid, can be mimicked by the trivial polarized state \cite{Chern2020}. 
In the absence of detailed calculations, we are presently not able to completely rule out the possibility that excitations in the 
trivial polarized state might also produce the charge redistribution around a vortex as discussed in our work.
Similarly, for the gapless case ($\kappa=0$) or for an Abelian variant of the gapped spin liquid, 
the electric polarization profile could potentially have a similar form.
However, the proposed absorption spectroscopy experiment (see Figs.~3 and 4 of the main text) will be able to clearly distinguish between the 
Kitaev spin liquid and other phases, since the MZM peak as well as the lowering of the continuum threshold
energy (by approximately a factor $1/2$) should only appear for the non-Abelian Kitaev spin liquid.  

We conclude by summarizing the robust aspects of our proposal for detecting $\mathbb{Z}_2$ vortices via their intrinsic 
electric field profile and, in particular, in absorption spectroscopy. When all other parameters are kept the same, the charge 
redistribution profile and the absorption spectrum (in the setup of Fig.~3 of the main text) must clearly change when a vortex is 
trapped or removed from a plaquette near an STM tip.  Our estimates for the corresponding voltage differences indicate that 
such tests are feasible with currently available experimental techniques.

\end{document}